%% file: main.tex
\documentclass[12pt, letterpaper]{article}

\usepackage[utf8]{inputenc}
\usepackage[english]{babel}
\usepackage{amsmath}
\usepackage{bbm}
\usepackage{amssymb}
\usepackage{mathtools}
\usepackage{amsthm}
\usepackage{fancyhdr}
\usepackage{titlesec}
\usepackage{graphicx}
\usepackage{csquotes}
\usepackage{multirow}
\usepackage{xcolor}
\usepackage{float}
\usepackage[round]{natbib}
\usepackage{changepage}
\usepackage{caption}
\usepackage{enumitem}
\usepackage{url}
\usepackage{lmodern}
\usepackage[T1]{fontenc}
\usepackage{xpatch}
\usepackage[linesnumbered,ruled,vlined]{algorithm2e}
\usepackage{ifthen}
\usepackage{listings}
\usepackage{subcaption}
\usepackage{chngcntr}
\usepackage{pdfpages}
\usepackage{natbib}
\usepackage{cases}
\usepackage{booktabs}
\usepackage[margin=1in]{geometry}

\graphicspath{{./images/} }

\begin{document}

\title{\bf Bayesian nonparametric Mallows model for clustering preference data}

\author{Lorenzo Zuccato$^{1,4}$\thanks{Corresponding author: \texttt{lorenzzu@math.uio.no}. This work was supported by the Research Council of Norway, Integreat – Norwegian Centre for Knowledge-driven Machine Learning, project number 332645, and by the Erasmus+ program of the European Union.} , Veronica Vinciotti$^2$, Valeria Vitelli$^{3,4}$\\
    \small{$1.$ Department of Mathematics, University of Oslo, Norway} \\
	\small{$2.$ Department of Mathematics, University of Trento, Italy} \\
    \small{$3.$ 
    Oslo Centre for Biostatistics and Epidemiology, University of Oslo, Norway} \\
    \small{$4.$ Integreat – Norwegian Centre for Knowledge-driven Machine Learning, Norway}\\
    }
\date{}
\maketitle

\begin{abstract}
Preference learning refers to the learning of latent patterns from ranking and preference data of different kinds. Typical aims of preference learning are to infer a shared consensus ranking, to learn individual-level preferences, and to perform unsupervised clustering. The Mallows model is among the few approaches that can achieve all these objectives jointly. Previous work has developed computationally tractable methods for Bayesian inference based on a MCMC Metropolis-Hastings scheme, where clustering is performed via a finite mixture of Mallows models. Inference on the number of clusters is then conducted a posteriori. Here we propose a Bayesian nonparametric Mallows model, based on a Dirichlet process mixture model. This allows joint inference on the number of non-empty clusters and on the clustering allocation, as well as posterior inference on cluster-specific parameters.
The implementation of the proposed  sampling algorithm is integrated into the existing \texttt{R} package \texttt{BayesMallows}, which also supports data in the form of incomplete rankings and pairwise comparisons. Simulated data show good performance of the nonparametric model compared to a finite mixture model in terms of recovery of the correct number of clusters, while empirical data on movie ratings show the model's effectiveness in providing personalized movie recommendations on discarded ratings. 
\end{abstract}


\section{Introduction}

Expressing opinions and preferences is a fundamental aspect of human nature, reflecting our cognitive, social, and decision-making processes. Whether done implicitly through behavior or explicitly through communication, the inclination to rank items is vital for self-expression. Preference learning has emerged as a subfield of statistics and machine learning focused on extracting insights from preference datasets of different kinds, such as rankings, ratings or pairwise comparisons collected from various assessors. Key tasks in this area include summarizing the different preferences into a consensus, estimating preferences in order to provide personalized recommendations, and clustering assessors based on their distinct preference patterns. However, the challenges introduced by the diversity and sparsity of such data have opened the way to the development of new methodologies.

While many methods are tailored to address specific challenges in preference learning, few demonstrate the flexibility to meet all key objectives simultaneously, while ensuring reliable uncertainty quantification. In this sense, the Bayesian version of the Mallows model, introduced by \citet{vitelli2018probabilistic}, stands out as a unified framework. This model is capable of handling partial rankings and pairwise preferences, managing various tasks while offering clear interpretability and broad applicability. In particular, the application of the method is not limited to the choice of analytically tractable distances, such as the Kendall distance, between an individual ranking and a consensus ranking.

Clustering users based on diverse preference patterns is particularly important when a high level of heterogeneity among assessors is present, which means that a single shared consensus is unrealistic. Moreover, partitioning data into groups with similar preferences leads to a better understanding of user behavior, thereby enhancing personalized recommendations. The Bayesian Mallows model of \citet{vitelli2018probabilistic} enables model-based clustering through finite mixtures, and uses a Markov Chain Monte Carlo (MCMC) Metropolis-within-Gibbs procedure for posterior sampling. However, model selection is challenging due to the lack of clear criteria for determining the number of clusters. Previous approaches have focused on the posterior distribution of within-cluster sum of distances, aiming to identify an elbow point indicative of the true number of clusters. Unfortunately, this often leads to inconclusive results, with multiple models appearing equally suitable or with an unclear elbow point, typically because of a linear decrease in the loss function.

In this paper, we address this challenge by considering Bayesian nonparametric methods. These methods provide a more flexible solution to clustering based on the overarching class of partition priors known as Gibbs-type priors, among which the Dirichlet process is probably the most established and widely used option. This approach, supported by efficient Metropolis-within-Gibbs sampling methods, can automatically accommodate an infinite number of clusters and therefore allows for fitting an infinite mixture of Mallows models. Notably, it automatically determines the number of non-empty clusters in the dataset while controlling complexity via the logarithmic clustering property of the Dirichlet process.

Previous work has already explored Bayesian nonparametric approaches for clustering ranking data. In particular, \citet{CaronBayesiannonp} and \citet{meilabayesiannonparametric} have employed the traditional Dirichlet process mixture for the Plackett-Luce and generalized Mallows models, respectively. However, the Plackett--Luce model \citep{Luce59, Plackett1975} can be restrictive in several respects. First, it assumes a sequential stagewise generation of rankings, which may be misspecified when the data do not arise from such a process and limits sensitivity to differences in lower-ranked positions. Second, being fully parameterized by item-specific scores, it does not operate directly on the permutation space, which can reduce the diversity of rankings it can represent, particularly beyond the top of the list. Finally, standard formulations do not include individual-level latent rankings, so inference and prediction are driven at the cluster level and are therefore inherently group-based. On the other hand, the generalized Mallows model is a distance-based model restricted to the use of few distances only, such as the Kendall \citep{fligner1986distance}, the Cayley \citep{irurozki2018sampling} and the Hamming \citep{irurozki2014sampling}, limiting both its applicability and flexibility. Other proposals \citep{meila2012estimation} have focused on non-parametric extensions of the generalized Mallows model for infinite sets of items, achieving clustering through the exponential blurring mean-shift approach.

Our proposed Dirichlet Process Mixture of Mallows Models (DPM3) combines interpretability and flexibility in exploring the permutation space, which is typical of the Mallows distribution, with the possibility of applying every right-invariant distance and the enhanced flexibility of the infinite mixture model in exploring the space of partitions. 
By evaluating the convergence of the chain, computing the probability that any two assessors coexist within the same cluster---the so-called co-clustering matrix---and establishing a partition of the data, we show how it is possible to obtain an a posteriori estimate of the number of non-empty clusters and to conduct inference on cluster-specific parameters. In particular, we discuss how conditioning on an estimated partition is essential in the case of an infinite mixture, since labels that appear and disappear throughout the algorithm's execution are not easily attributable to the same underlying cluster, as instead is the case with label switching reordering algorithms for finite mixtures. 

In Section~\ref{sec:DPM3}, we present the Dirichlet process mixture of Mallows models, including model specification and a description of the sampling algorithms for both complete and incomplete data. This section also presents techniques for deriving meaningful posterior inference. Section~\ref{sec:simulation} provides a simulation study to evaluate the performance of our approach in recovering the correct number of non-empty clusters and detecting missing pairwise preferences, in comparison to a finite mixture model. Finally, Section~\ref{sec:movieratings} provides an illustration of the method to the case of movie ratings showing how the model is able to provide personalized movie recommendations on discarded ratings.

\section{Dirichlet process mixture of Mallows models} \label{sec:DPM3}

In this section, we present our proposal to integrate a Bayesian nonparametric approach within the Bayesian Mallows model framework of \citet{vitelli2018probabilistic}. In particular, we use the Dirichlet process mixture model \citep{Antoniak73, Ferguson83}. This model is designed to non-parametrically combine parametric families, enabling the clustering of data into a potentially infinite number of components. For this reason, it is also known as infinite mixture model. Advances in MCMC sampling algorithms \citep{Escobar1995, Mac1998} have made the Dirichlet process mixture model feasible by enabling efficient inference on the posterior distribution of parameters. For a summary of these algorithms, see \citet{neal2000}.

\subsection{Finite mixture of Mallows models}\label{sec:prelim}

We consider the following modeling setting. Assume that $N$ assessors are asked to rank a set of $n$ labeled items $\mathcal{A}=\{A_1,\dots,A_n\}$. Then, a full ranking can be seen as a mapping $\mathbf{R}:\mathcal{A}\rightarrow \mathcal{P}_n$, where $\mathcal{P}_n$ is the space of $n$-dimensional permutations. We denote with $\mathbf{R}_1,\dots,\mathbf{R}_N$ the rankings of the assessors, where we interpret $R_{ik}<R_{il}$ as item $A_k$ preferred to item $A_l$ by assessor $i$, and we say in this case that $A_k$ is ranked higher than $A_l$. The Mallows model assumes these rankings to be generated by a probability distribution over the space $\mathcal{P}_n$. Specifically:
\begin{equation}
    \label{eqn:MM}
    P(\mathbf{R}_1,\ldots,\mathbf{R}_N \mid \alpha, \boldsymbol{\rho}) = \frac{1}{Z_n(\alpha)^N}\exp\left\{-\frac{\alpha}{n}\sum_{j=1}^N d(\mathbf{R}_j,\boldsymbol \rho) \right\}\prod_{j=1}^N\mathbbm{1}_{\mathcal{P}_n}(\mathbf{R}_j),
\end{equation}
where $\boldsymbol{\rho} \in \mathcal{P}_n$ is the location parameter, also known as consensus ranking, $\alpha\ge 0$ is the scale parameter, also known as precision, $d(\cdot,\cdot)$ is a right-invariant distance measure between permutations, and $Z_n(\alpha)$ is the normalizing constant, which depends only on $\alpha$ due to the right-invariance of the distance.

The combination of model \eqref{eqn:MM} with a finite mixture has been used to capture heterogeneity among the assessors, grouping them into different clusters. In this case, the ranks are assumed to be generated from a mixture of Mallows distributions with cluster-specific parameters $\alpha_{c}$ and $\boldsymbol \rho_{c}$. A Dirichlet prior for the cluster proportions and a multinomial distribution on the cluster labels complete the hierarchical model:
\begin{align}
\begin{split}
    \mathbf{R}_j \mid \{\alpha_c, \boldsymbol \rho_c\}_{c=1}^C, z_j \; &\sim \; \text{Mallows}(\alpha_{z_j}, \boldsymbol \rho_{z_j}) \\
    (\alpha_c, \boldsymbol \rho_c)\; &\sim\; \text{TruncExp}(\lambda, \alpha_{\max})\times \text{Unif}(\mathcal{P}_n)\\
    z_j \mid \tau_1,\dots,\tau_C\; &\sim\; \text{Mult}(\tau_1,\dots,\tau_C) \\
    (\tau_1,\dots,\tau_C)\; &\sim\; \text{Dir}(\psi,\dots,\psi).
\end{split}
\end{align}
The number of components $C$ is fixed, and various ways can be employed to determine its most appropriate value. In particular, \citet{vitelli2018probabilistic} propose the inspection of the posterior distributions of the within-cluster sum of distances over a number of different choices of $C$.

\subsection{Infinite mixture of Mallows models}

In order to specify the Dirichlet Process Mixture of Mallows Models (DPM3), we make use of the Dirichlet process mixture model in its infinite mixture formulation. 
Using the notation introduced in Section \ref{sec:prelim}, the model is defined by
\begin{equation}
\label{eqn:ourmodel}
    \begin{split}
        \mathbf{R}_i \mid z_i, \boldsymbol{\theta}^* \; &\sim \; \text{Mallows}( \alpha_{z_i}, \boldsymbol \rho_{z_i}) \\
        z_i\mid \boldsymbol \tau \; &\sim \; \text{Mult}(\tau_1,\dots,\tau_C) \\
        (\alpha_c, \boldsymbol \rho_c) \; &\sim \; G_0 \equiv \text{TruncExp}(\lambda, \alpha_{\max})\times \text{Unif}(\mathcal{P}_n)\\
        \boldsymbol \tau \; &\sim \; \text{Dir}(\psi/C, \dots, \psi/C),
    \end{split}
\end{equation}
where $i=1,\dots,N$, $c=1,\dots,C$ and $\boldsymbol\theta ^*=((\alpha_1, \boldsymbol \rho_1),\dots,(\alpha_C,\boldsymbol \rho_C))$ and the obvious conditional independence properties are tacitly assumed. The mixing proportions are assigned a symmetric Dirichlet prior distribution whose concentration parameter goes to 0 as $C$ goes to infinity. Indeed, it has been shown that, when $C$ goes to infinity, this formulation is equivalent to the assignment of a Dirichlet process prior with concentration parameter $\psi$ and base measure $G_0$ on the distribution of the clusters' parameters \citep{neal2000}. Also, notice that we are assuming truncated exponential priors for the precision parameters $\alpha_c$ and uniform priors for the consensus rankings $\boldsymbol{\rho}_c$.

Using known results on the full conditional distributions of the latent allocation variables, we obtain that, for $C \to \infty$, the conditional probability of $z_i$ given all other allocations, the observed data and the clusters' parameters can be written as
\begin{subnumcases}{P(z_i = c \mid \mathbf{z}^{(-i)}, \mathbf{R}_i, \boldsymbol{\theta}^*) =}
\label{eqn:ourprobexistingcluster}
\frac{b\,N_{c,\mathbf{z}^{(-i)}}}{\psi+N-1}
P(\mathbf{R}_i\mid \alpha_c, \boldsymbol \rho_c)
\qquad \qquad  \text{if } c = z_j \text{ for some } j \neq i &
\\
\label{eqn:ourprobnewcluster}
\frac{b\,\psi}{\psi+N-1}
\int_{\Theta} P(\mathbf{R}_i\mid \alpha, \boldsymbol{\rho})\,\mathrm{d}G_0(\alpha,\boldsymbol \rho)
\qquad \text{if }  z_i \neq z_j \ \forall j \neq i &
\end{subnumcases}
where $N_{c,\mathbf{z}^{(-i)}} = \sum_{j\neq i} \mathbbm{1}(z_j=c)$ denotes the number of observations allocated to cluster $c$, excluding observation $i$. The parameters $\alpha$ and $\boldsymbol{\rho}$ denote cluster-specific parameters with joint parameter space $\Theta = \mathcal{P}_n \times [0, \alpha_{\max}]$, and $G_0$ is a bivariate prior distribution defined on $\Theta$. The constant $b$ is a normalizing term ensuring that the conditional probabilities sum to one over all possible values of $z_i$, including both existing and new clusters. These expressions define the full conditional distribution of the allocation variables $z_i$, and they provide the key ingredient for constructing a sampling algorithm for updating cluster assignments.

In order for the sampling algorithm to be useful in practice, we need to be able to compute the integral in Equation \eqref{eqn:ourprobnewcluster}. Since the priors for $\alpha$ and $\boldsymbol \rho$ are independent, the joint prior is $G_0(\alpha,\boldsymbol \rho)= \pi(\alpha)\pi(\boldsymbol \rho)$, where 
\begin{align*}
\pi(\alpha)
= \frac{\lambda \exp\{-\lambda \alpha\}}{1-\exp\{-\lambda\alpha_{\max}\}}
\mathbbm{1}_{[0,\alpha_{\max}]}(\alpha),
\hspace{2cm}
\pi(\boldsymbol{\rho})
&= \frac{1}{n!}\mathbbm{1}_{\mathcal{P}_n}(\boldsymbol \rho).
\end{align*}
Therefore:
\begin{align*}
\int_\Theta P(\mathbf{R}_i \mid \alpha, \boldsymbol{\rho}) \, \mathrm{d}G_0(\alpha,\boldsymbol{\rho})
&= \int_0^{\alpha_{\max}}
\frac{1}{n!}
\sum_{\mathbf{r}\in \mathcal{P}_n}
\frac{1}{Z_n(\alpha)}
\exp \left\{-\frac{\alpha}{n} d(\mathbf{R}_i, \mathbf{r})\right\}
\frac{\lambda e^{-\lambda \alpha}}{1-e^{-\lambda \alpha_{\max}}}
\, \mathrm{d}\alpha \\
&= \frac{1}{n!}
\int_0^{\alpha_{\max}}
\frac{\lambda e^{-\lambda \alpha}}{1-e^{-\lambda \alpha_{\max}}}
\, \mathrm{d}\alpha \\ &= \frac{1}{n!}.
\end{align*}
where we used the fact that, due to the right-invariance of $d(\cdot, \cdot),$
\[
\sum_{\mathbf{r}\in \mathcal{P}_n}\exp \left\{-\frac{\alpha}{n}d(\mathbf{R}_i, \mathbf{r})\right\}=
\sum_{\mathbf{r}\in \mathcal{P}_n}\exp \left\{-\frac{\alpha}{n}d(\mathbf{r}, \mathbf{e})\right\}=Z_n(\alpha),
\]
for any reference ranking $\mathbf{e}\in \mathcal{P}_n.$

Moreover, when initializing the parameters of a newly generated cluster, it is convenient to sample them from the distribution $H_i(\alpha, \boldsymbol \rho)$, which is the posterior of the parameters given the single observation $\mathbf{R}_i$. This is given by
\begin{align*}
H_i(\alpha,\boldsymbol{\rho})
&= \frac{P(\mathbf{R}_i\mid \alpha,\boldsymbol{\rho})\,G_0(\alpha,\boldsymbol{\rho})}
{\int_\Theta P(\mathbf{R}_i\mid \alpha, \boldsymbol{\rho})\,\mathrm{d}G_0(\alpha,\boldsymbol{\rho})} \\
&= \frac{1}{Z_n(\alpha)}
\exp\left\{-\frac{\alpha}{n}d(\mathbf{R}_i, \boldsymbol{\rho}) \right\}
\mathbbm{1}_{\mathcal{P}_n}(\boldsymbol{\rho})
\frac{\lambda e^{-\lambda \alpha}}{1-e^{-\lambda \alpha_{\max}}}
\mathbbm{1}_{[0,\alpha_{\max}]}(\alpha) \\
&= \pi(\boldsymbol{\rho} \mid \mathbf{R}_i, \alpha)\,\pi(\alpha).
\end{align*}
The solution is to first sample $\alpha$ from the truncated exponential distribution, and then sample $\boldsymbol \rho$ from a Mallows density with location and precision parameters $\mathbf{R}_i$ and $\alpha$, respectively. More details on the implementation of this step and the sampling algorithm are given in the next sections.

\subsection{MCMC algorithm for sampling from the DPM3 posterior}

The posterior distribution of the DPM3 model is explored using a Markov chain Monte Carlo (MCMC) algorithm based on a Metropolis-within-Gibbs scheme, which allows sampling from the model in Equation \eqref{eqn:ourmodel} in the limit as $C \rightarrow \infty$. The algorithm combines updates of the cluster allocation variables with updates of model parameters, where the former are driven by the full conditional distributions derived in the previous section, and the latter are handled through Metropolis-type steps. To simplify the presentation, we first consider the case of complete rankings and describe the corresponding MCMC scheme, illustrated in Algorithm~\ref{eq:algorithm1}.
\begin{algorithm}
    \small
    \SetAlgoLined
    \caption{DPM3 MCMC Algorithm for Complete Rankings}
    \LinesNotNumbered
    \SetKwInOut{Input}{input}\SetKwInOut{Output}{output}

    \Input{$\mathbf{R}_1,\dots,\mathbf{R}_N$; $\psi$, $\psi_{\text{init}}$, $\lambda$, $\sigma_{\alpha}$, $\alpha_{\text{jump}}$, $L$, $d(\cdot,\cdot)$, $Z_n(\alpha)$, $M$.}
    \Output{Posterior distributions of $z_1,\dots,z_N$, $\{\boldsymbol \rho_c\}_{c=1}^\infty$, $\{\alpha_c\}_{c=1}^\infty$.}
    \vspace{3pt}
    \textbf{Initialization of the MCMC:}  
    \begin{itemize}
    \setlength\itemsep{-5pt}
        \item sample: $z_{1,0},\dots,z_{N,0} \sim \text{CRP}(\psi_{\text{init}})$
        \item compute: $C_0 \leftarrow $ unique$\{z_{1,0},\dots,z_{N,0}\}$
        \item  set $\alpha_{1,0}= \dots=\alpha_{C_0, 0}=1$ and generate $\boldsymbol \rho_{1,0},\dots,\boldsymbol \rho_{C_0, 0} \sim \text{Unif}(\mathcal{P}_n)$ 
    \end{itemize}

    \For{$m \leftarrow 1$  \KwTo $M$}{
        \vspace{4pt}
        \textbf{Step 1: update clusters parameters} \\
            \For{$c$ \textnormal{\textbf{in}} \textnormal{unique}$\{\mathbf{z}_{m-1}\}$}{
                \textbf{M-H step: update }$\boldsymbol \rho_c$ \\
                sample: $\boldsymbol \rho_c' \sim \text{L\&S}(\boldsymbol \rho_{c,m-1}, L)$ and  $u \sim \text{Unif}(0,1)$ \\
                compute: $ratio \leftarrow$ equation \eqref{eqn:ratiorho} with $\boldsymbol \rho \leftarrow \boldsymbol \rho_{c,m-1}$ and $\alpha \leftarrow \alpha_{c,m-1}$ \\
                \textbf{if} $u<ratio$ \textbf{then} $\boldsymbol \rho_{c,m} \leftarrow \boldsymbol \rho_c'$, \textbf{else} $\boldsymbol \rho_{c,m} \leftarrow \boldsymbol \rho_{c,m-1}$\\
                \vspace{10pt}
                \textbf{if } $m \mod \alpha_{\text{jump}}=0$ \textbf{then M-H step: update} $\alpha_c$ \\
                
                sample: $\alpha_c' \sim \log N(\log(\alpha_{c,m-1}), \sigma^2_\alpha$) and  $u \sim \text{Unif}(0,1)$ \\
                compute: $ratio \leftarrow$ equation \eqref{eqn:ratioalpha} with $\boldsymbol \rho \leftarrow \boldsymbol \rho_{c,m}$ and $\alpha \leftarrow \alpha_{c,m-1}$ and where the sum is over $\{j:z_{j,m-1}=c\}$\\
                \textbf{if} $u<ratio$ \textbf{then} $\alpha_{c,m} \leftarrow \alpha_c'$, \textbf{else} $\alpha_{c,m} \leftarrow \alpha_{c,m-1}$\\
            }
        \vspace{4pt}
        \textbf{Step 2: update cluster {assignments}} \\
            \For{$i\leftarrow 1 $ \KwTo $N$}{
                \textbf{Gibbs step: update }$z_i$\\
                \textbf{if} $z_{i,m-1}\neq z_{j,m}$ {for} $j=1,\dots,i-1$ and $z_{i,m-1}\neq z_{j,m-1}$ for $j=i+1,\dots,N$ \textbf{then} remove $\boldsymbol \rho_{z_{i,m-1},m-1}$ and $\alpha_{z_{i,m-1},m-1}$ from the state
            
            \For{$c$ \textnormal{\textbf{in} unique}$\{z_{1,m},\dots,z_{i-1,m},z_{i+1,m-1},\dots,z_{N,m-1}\}$}{
                ${p}_{i,c} \leftarrow$ equation \eqref{eqn:ourprobexistingcluster} with $n_{c,\mathbf{z}^{(-i)}}\leftarrow$ equation \eqref{eqn:countingforprob}
            }
            ${p}_{i,\text{new}}\leftarrow $ equation \eqref{eqn:ourprobnewcluster} \\
            sample: $z_{i,m} \sim \text{Mult}((p_{i,c})_c, {p}_{i,\text{new}})$ \\
            \If{ $z_{i,m}\neq z_{j,m}$ \textnormal{for} $j=1,\dots,i-1$ \textnormal{and} $z_{i,m}\neq z_{j,m-1}$ \textnormal{for} $j=i+1,\dots,N$}{
            sample: $\alpha_{z_{i,m},m} \sim \text{TruncExp}(\lambda,\alpha_{\max})$\\
            sample: $\boldsymbol \rho_{z_{i,m},m}\sim\text{Mallows}(\mathbf{R}_i,\alpha_{z_{i,m},m})$
            }
            }
    }
\label{eq:algorithm1}
\end{algorithm}

Since the number of non-empty clusters is unknown, in order to guarantee a random and coherent initialization of the labels, we use a Chinese restaurant process with parameter $\psi_{\text{init}}$. Tuning this parameter does not influence the algorithm convergence. Once the number of initial clusters $C_0$ is set, we can randomly generate $\boldsymbol \rho_{1,0},\dots,\boldsymbol \rho_{C_0,0}$ and set $\alpha_{1,0},\dots,\alpha_{C_0,0}$ to be equal to 1 for simplicity.

The updates for the cluster-specific parameters $\boldsymbol \rho$ and $\alpha$ exclusively involve clusters that are currently non-empty, and employ the Metropolis-Hastings technique, using the following ratios presented in \citet{vitelli2018probabilistic}:
\begin{equation}
    \label{eqn:ratiorho}
    \min\left\{1, \frac{P_L(\boldsymbol \rho \mid \boldsymbol \rho')}{P_L(\boldsymbol \rho' \mid \boldsymbol \rho)} \exp \left[-\frac{\alpha}{n}\sum_{j=1}^N\{d(\mathbf{R}_j, \boldsymbol \rho')-d(\mathbf{R}_j, \boldsymbol \rho)\}\right]\right\},
\end{equation}
and 
\begin{equation}
    \label{eqn:ratioalpha}
    \min\left\{1,\frac{Z_n(\alpha)^N\pi(\alpha')\alpha'}{Z_n(\alpha')^N\pi(\alpha)\alpha}\exp\left[-\frac{(\alpha'-\alpha)}{n}\sum_{j=1}^Nd(\mathbf{R}_j, \boldsymbol \rho)\right]\right\}.
\end{equation}
In the acceptance probabilities above, we employ the proposal distributions used in the original formulation of the Bayesian Mallows model. In particular, $\boldsymbol{\rho}'$ is proposed according to a Leap-and-Shift (L\&S) distribution $P_L(\boldsymbol{\rho}' \mid \boldsymbol{\rho})$, with tuning parameter $L \in \{1,\dots,\lfloor (n-1)/2 \rfloor\}$, as defined in \citet{vitelli2018probabilistic}. For the scalar parameter $\alpha$, we use a log-normal proposal of the form $\alpha' \sim \log\mathcal{N}(\log \alpha, \sigma_\alpha^2)$.

Throughout the algorithm we use the index $m$ to keep track of the current iteration. As the Gibbs stage for the assignment of labels has the potential to create a new cluster for each assessor, we must compute the probabilities separately while scanning through all the assessors. In the process of updating the assignment for assessor $i$ at iteration $m$, assessors $1,\dots,i-1$ have already undergone updates to $z_{1,m},\dots,z_{i-1,m}$, whereas assessors $i+1,\dots,N$ are still assigned based on $z_{i+1,m-1},\dots,z_{N,m-1}$. Therefore, while computing the probabilities in Equation \eqref{eqn:ourprobexistingcluster}, the counting of the cardinality of groups must be based on $z_{1,m},\dots,z_{i-1,m},z_{i+1,m-1},\dots,z_{N,m-1}$, as follows:
\begin{equation}
    \label{eqn:countingforprob}
    N_{c,\mathbf{z}^{(-i)},m} = \sum_{j=1}^{i-1}\mathbbm{1}(z_{j,m}=c)+\sum_{j=i+1}^N\mathbbm{1}(z_{j,m-1}=c).
\end{equation}
These assignment probabilities can be computed up to a normalizing constant $b$, but this does not affect the implementation.

When a new cluster is randomly drawn, initializing its parameters involves sampling from the previously derived $H_i(\alpha,\boldsymbol \rho)$.  As sampling from a Mallows distribution is generally a complex task, we leverage the MCMC method proposed in Algorithm 5 of \citet{vitelli2018probabilistic}. Since we only need to generate one ranking, the thinning parameter is unnecessary, while the burn-in is chosen to be proportional to the logarithm of the number of items $n$ to ensure an effective convergence.

Note that the parameter $\psi$ controls the probability of assigning an assessor to a new empty cluster at each Gibbs step. A higher value of $\psi$ encourages exploration of configurations with more clusters, helping to avoid convergence to solutions with minimal clustering structure, at the cost of increased computational demands.

In many applications, the available data consist only of partial rankings or pairwise comparisons rather than complete rankings. The DPM3 model can be naturally extended to accommodate such incomplete information by incorporating data augmentation steps analogous to those introduced for the Bayesian Mallows model. The idea is to introduce augmented rankings $\Tilde{\mathbf{R}}_1,\dots,\Tilde{\mathbf{R}}_N$ that are compatible with the limited information contained in the partial rankings $\mathbf{R}_1, \dots, \mathbf{R}_N$ or the sets of transitive pairwise preferences $\mathcal{B}_1,\dots,\mathcal{B}_N$. 
At each iteration $m$, the augmented rankings are updated via a Metropolis--Hastings step. For each $j$, a candidate ranking $\tilde{\mathbf{R}}_j'$ is proposed either by sampling uniformly from the set of rankings compatible with the observed partial ranking $\mathbf{R}_j$, or via a modified L\&S move respecting the pairwise comparisons $\mathcal{B}_j$; both proposals are symmetric. The acceptance probability thus reduces to
\begin{equation}
    \label{eqn:ratioaugmented}
    \min\left\{ 1, \exp \left[ -\frac{\alpha}{n} \left( d(\Tilde{\mathbf{R}}_j', \boldsymbol \rho)-d(\Tilde{\mathbf{R}}_j, \boldsymbol \rho)\right) \right] \right\}
\end{equation}
 The corresponding DPM3 sampling procedure is outlined in Algorithm~\ref{eq:algorithm2}.
\begin{algorithm}
    \small
    \SetAlgoLined
    \caption{DPM3 MCMC Algorithm for Incomplete Data}
    \LinesNotNumbered
    \SetKwInOut{Input}{input}\SetKwInOut{Output}{output}

    \Input{$\{\mathcal{S}_1,\dots,\mathcal{S}_N\}$ or $\{\text{tc}(\mathcal{B}_1),\dots,\text{tc}(\mathcal{B}_N)\}$; $\psi$, $\psi_{\text{init}}$, $\lambda$, $\sigma_{\alpha}$, $\alpha_{\text{jump}}$, $L$, $d(\cdot,\cdot)$, $Z_n(\alpha)$, $M$.}
    \Output{Posterior distributions of $z_1,\dots,z_N$, $\Tilde{\mathbf{R}}_1,\dots,\Tilde{\mathbf{R}}_N$, $\{\boldsymbol \rho_c\}_{c=1}^\infty$, $\{\alpha_c\}_{c=1}^\infty$.}
    \vspace{3pt}
    \textbf{Initialization of the MCMC:} \\
        generate: $z_{1,0},\dots,z_{N,0}$, $\alpha_{1,0}, \dots,\alpha_{C_0, 0}$, $\boldsymbol \rho_{1,0},\dots,\boldsymbol \rho_{C_0, 0}$ as in algorithm 1 \\
        
    \uIf{$\{\mathcal{S}_1,\dots,\mathcal{S}_N\}$ among inputs}{\textbf{for} $j\leftarrow 1$ \textbf{to} $N$ \textbf{do} randomly generate $\Tilde{\mathbf{R}}_{j,0}$ in $\mathcal{S}_j$}
    \Else{
    \textbf{for} $j\leftarrow 1$ \textbf{to} $N$ \textbf{do} randomly generate $\Tilde{\mathbf{R}}_{j,0}$ compatible with tc$(\mathcal{B}_j)$
    }
    
    \vspace{10pt}
      \For{$m \leftarrow 1$  \KwTo $M$}{
        \vspace{4pt}
        \textbf{Step 1: update clusters parameters} \\
            as in algorithm 1 with $\mathbf{R}_j \leftarrow \Tilde{\mathbf{R}}_{j,m-1}$ \\
        \vspace{4pt}
        \textbf{Step 2: update cluster {assignments}} \\
            as in algorithm 1 with $\mathbf{R}_j \leftarrow \Tilde{\mathbf{R}}_{j,m-1}$ \\
        \vspace{4pt}
        \textbf{Step 3: update augmented rankings} \\
        \For{$j\leftarrow 1$ \KwTo $N$}{
            \textbf{M-H step: update $\Tilde{\mathbf{R}}_j$} \\
            \uIf{$\{\mathcal{S}_1,\dots,\mathcal{S}_N\}$ among inputs}{
            sample: $\Tilde{\mathbf{R}}_j'$ uniformly in $\mathcal{S}_j$ 
            }
            \Else{
            sample: $\Tilde{\mathbf{R}}_j'$ compatible with tc$(\mathcal{B}_j)$ from $\text{L\&S}(\Tilde{\mathbf{R}}_{j,m-1}, L)$
            }
            compute: $ratio \leftarrow$ equation \eqref{eqn:ratioaugmented} with $\boldsymbol \rho \leftarrow \boldsymbol \rho_{z_{j,m},m}$, $\alpha \leftarrow \alpha_{z_{j,m},m}$ and $\Tilde{\mathbf{R}}_j \leftarrow \Tilde{\mathbf{R}}_{j,m-1}$ \\
            sample: $u\sim\text{Unif}(0,1)$ \\
            \textbf{if} $u<ratio$ \textbf{then} $\Tilde{\mathbf{R}}_{j,m}\leftarrow \Tilde{\mathbf{R}}_j'$ \textbf{else} $\Tilde{\mathbf{R}}_{j,m} \leftarrow \Tilde{\mathbf{R}}_{j,m-1}$
        }
    }
 \label{eq:algorithm2}   
\end{algorithm}

\subsection{Posterior inference for the parameters of DPM3}
After obtaining samples of cluster assignments and cluster-specific parameters with appropriate thinning,  posterior inference can be conducted and is discussed in this section. For simplicity of notation, we assume without loss of generality that both the thinning parameters and $\alpha_{\text{jump}}$ are set to 1, ensuring that there is one sample for every iteration $m$.

The posterior distribution of the number of non-empty clusters is derived by recording, at each iteration, the count of unique labels among $z_{1,m}, \dots, z_{N,m}$. Unfortunately, this distribution frequently leads to an overestimation of the true number of non-empty clusters, especially when the parameter $\psi$, which governs the probability of assignment to an empty cluster, is large \citep{miller2013simple, yang2019posterior}. For this reason, it is crucial to verify the persistence of the non-empty clusters throughout the chain. The greater the number of iterations in which a particular cluster is persisting, the more likely it is that such cluster is non-empty.

It is important to note that the use of a Dirichlet process mixture approach, as compared to finite mixture modeling, leads to a loss of cluster identifiability. Specifically, cluster labels are assigned as progressive natural numbers for newly generated clusters, implying that, when a cluster becomes empty, its label is not reassigned. The identifiability challenge thus arises from the inability to discern which distinct labels correspond to the same cluster becoming empty and reappearing later along the chains. Furthermore, label switching may occasionally occur, and addressing it offline in the context of an infinite mixture can be troublesome. For these reasons, it is common practice to interpret the result of the chain as a posterior distribution over the space of partitions of the $N$ data points, and to derive summaries of this distribution \citep{medvedovic2002bayesian, dahl2006model, rajkowski2016analysis, WadeBayesian}. Subsequently, cluster-specific parameter inference can be performed after conditioning on the specific estimated partition of the data. This highlights a fundamental difference in posterior inference between finite mixture models and Dirichlet process mixture models. In finite mixture models, offline correction for label switching is typically sufficient to recover coherent cluster assignments and to perform cluster-specific posterior inference.
In contrast, Dirichlet process mixture models require a more intricate approach, due to the inherent uncertainty in the number of clusters and their identities.

The summary statistic most commonly used to capture the underlying clustering structure is the co-clustering matrix, here denoted with $\Xi \in [0,1]^{N\times N}$. Each cell of the co-clustering matrix, $\xi_{ik}:=\{\Xi\}_{ik},$ for $i,k=1,\ldots,N,$ reports the probability that assessors $i$ and $k$ coexist within the same cluster in the posterior samples, that is
\begin{equation}
    \label{eqn:co-clus}
    \xi_{ik}= \frac{1}{M-B}\sum_{m=B+1}^M \mathbbm{1}(z_{i,m}=z_{k,m}),
\end{equation}
where $B$ represents an appropriate number of iterations to discard as a burn-in.
These entries clearly form a symmetric matrix. 

Conditioning upon the co-clustering matrix allows for a straightforward estimation of the number of non-empty clusters in the data, $\hat{C}$, along with the corresponding data partition $\hat{z}_1,\dots,\hat{z}_N \in \{1,\dots, \hat{C}\}$. A number of approaches can be used to this aim. One of the most rigorous methods was developed by \citet{WadeBayesian}, and it involves  optimizing the lower bound to the posterior expected variation of information \citep{meilua2007comparing}, as determined by Jensen’s inequality. To enhance computational efficiency, a modified posterior expected variation of information is used, where the logarithm and expectation are exchanged. This modification speeds up computing time significantly, relying solely on the posterior co-clustering matrix for computational dependencies (for more details see \citealp{mcclustext}). Other possible loss functions have been proposed \citep{fritsch2009improved,lau2007bayesian, quintana2003bayesian}. Alternatively, heuristic approaches such as K-means and hierarchical clustering methods can also extract valuable insights from the co-clustering matrix, aiding in the final decision regarding the number of sub-groups and the partition. Notice that the final values estimated in the variables $\hat{z}_1,\dots,\hat{z}_N$ do not necessarily correspond to the labels initially assigned by the MCMC algorithm. 

By conditioning on the estimated partition, we can then draw inferences regarding cluster-specific parameters $\{\boldsymbol \rho^*_c \}_{c=1}^{\hat{C}}$ and $\{\alpha^*_c \}_{c=1}^{\hat{C}}$. This involves interpreting samples $\boldsymbol \rho_{z_{i,m},m}$ and $\alpha_{z_{i,m},m}$, where $m=1,\dots,M$, and for all $i=1,\ldots,N$ such that $\hat{z}_{i}=c$, as samples from the posterior distribution of $\boldsymbol \rho^*_{c}$ and $\alpha_c^*$. As a result, it becomes feasible to derive posterior summaries for the parameters, including means, medians, and credible intervals. This interpretation implies that, in the posterior distribution, the same parameter values are repeated a number of times equal to the number of assessors assigned to the corresponding cluster at the given iteration. This redundancy compensates for the influence of iterations in which assessors that are grouped together in the estimated partition are instead assigned to very different clusters, thereby resulting in highly dissimilar associated parameters.

The estimation of posterior summaries for the consensus ranking, which is defined in a permutation space, deserves further attention. Computing the maximum a posteriori (MAP) of the consensus is not only computationally challenging, but also not particularly suitable from a modeling perspective due to the discreteness and huge cardinality of the space $\mathcal{P}_n$. Consequently, we propose to adopt the methodology introduced by \citet{vitelli2018probabilistic}, which involves computing the cumulative probability consensus ranking. This consensus is established through a sequential process: initially selecting the item with the maximum a posteriori marginal probability of being ranked $1^{\text{st}}$, then choosing the item with the maximum a posteriori marginal probability of being ranked $1^{\text{st}}$ or $2^{\text{nd}}$ among the remaining items, and so forth. This procedure allows to leverage the major advantage of the Bayesian framework, that is uncertainty quantification. Similar techniques can be used to estimate other functions of the cluster-specific consensus rankings or of the augmented individual rankings. 

\subsection{Implementation of DPM3}
We implemented the algorithm by extending the \texttt{R} package \texttt{BayesMallows} \citep{BayesMallowsRjournal}, which allowed us to exploit the flexibility and versatility of the Bayesian Mallows model in an efficient way. Specifically, we wrote the core MCMC algorithm in \texttt{C++}, made executable via the \texttt{R} function \texttt{compute\_mallows\_dpmixture} through the \texttt{RCpp} package \citep{rcpp}. This returns objects of class \texttt{BayesMallowsDPMixture}, with associated functionalities to visualize chain convergence, compute the co-clustering matrix, and conduct posterior inference, as outlined in the previous subsection. The code is publicly available on a GitHub repository\footnote{\url{https://github.com/lorenzo-zuccato/BayesMallows-dpmixture/tree/dpmixture}} forked from the original \texttt{BayesMallows} package. For further details on the use of the package we refer to the Supplementary Material.

\section{Experiments on simulated data} \label{sec:simulation}

\begin{table}
    \centering
    \caption{Cluster-specific sizes, consensus rankings, and precision parameters used in the generation of the \textit{top8} and \textit{pref30} simulated datasets.}
    \label{tab:settingtoppref}
    \begin{tabular}{lccc}
        \toprule
        Cluster & Size & $\boldsymbol{\rho}$ & $\alpha$ \\
        \midrule
        1 & 50 & $(1,\dots,30)$          & 3 \\
        2 & 25 & $(30,\dots,1)$          & 2 \\
        3 & 25 & $(16,\dots,30,1,\dots,15)$ & 5 \\
        \bottomrule
    \end{tabular}
\end{table}

In this section, we evaluate the proposed model and inferential procedure on synthetic datasets generated from a finite mixture of Mallows models. We compare the ability of DPM3 to recover the correct number of non-empty clusters with that of a finite mixture approach. To this end, we simulate datasets using the procedure outlined in Algorithm 5 of \citet{vitelli2018probabilistic}. We focus on two scenarios that are particularly relevant in applications, namely top-$k$ rankings and pairwise preferences. In particular, we consider two levels of sparsity, denoted \textit{top8} and \textit{pref30}, respectively. For both scenarios, we generate 30 datasets with $N=100$ assessors and $n=30$ items. Each dataset is generated from a three-component mixture model with cluster sizes 50, 25, and 25, and component-specific parameters as reported in Table \ref{tab:settingtoppref}.

Having sampled full rankings for each assessor $\mathbf{R}_1, \dots, \mathbf{R}_{100}$, we then apply a procedure to sparsify the dataset. To obtain the \textit{top8} datasets we retain information only on the first $k_i$ ranked items for each assessor $i=1,\ldots,N$, where $k_i$ is sampled from a Poisson distribution with mean 8, truncated at 30. On the other hand, for the \textit{pref30} scenario, we sample $k_i \sim \text{Pois}(30)$ random pairs of objects $(A_j, A_k)$ with $j,k = 1,\ldots,n, \ j \neq k$ and without replacement, storing the corresponding comparison according to the true latent ranking $\mathbf{R}_i$ in the set $\mathcal{B}_i$. This generative process always yields transitive preferences as its outcome.

In both scenarios, we conduct analyses using finite and infinite mixture models, setting both $\alpha_{\text{jump}}$ and the thinning parameter to $10$. For the finite mixture model, we consider $C = 2, \dots, 6$ and run $5 \times 10^4$ iterations, storing all required within-cluster sums of distances. For the nonparametric counterpart, we fix $\psi = 0.025$ and similarly run $5 \times 10^4$ iterations. In all cases, this number of iterations proved sufficient for the Markov chain to converge to a stable outcome.

Upon convergence of the algorithms, we evaluate the ability of each method to accurately identify the true number of clusters across all simulated datasets. In the case of finite mixture models, we discard $2 \times 10^4$ iterations for each model as burn-in and visualize the posterior distributions of within-cluster distances over the remaining samples for each dataset. For the DPM3 model, the burn-in period is determined separately for each experiment based on inspection of the trace plots of $\alpha$, reflecting the greater variability in convergence behavior induced by the larger partition space explored by the Dirichlet process mixture. The results are then summarized through co-clustering matrices, which are subsequently used to estimate a partition and, consequently, the corresponding number of non-empty clusters, as described in Section~\ref{sec:DPM3}.

\begin{figure}[t]
    \centering
    \begin{subfigure}{0.325\textwidth}
        \includegraphics[width=\linewidth]{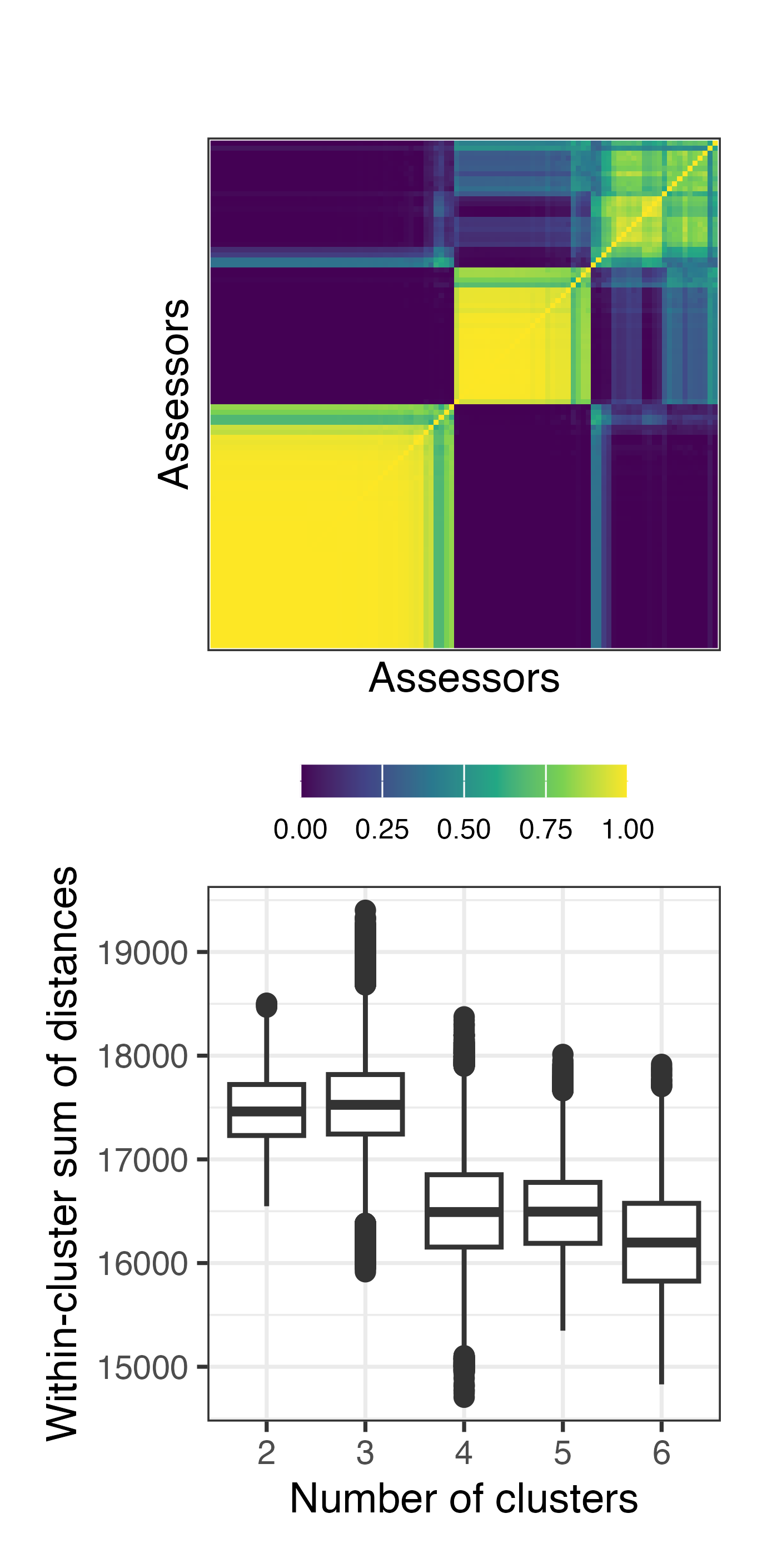}
        \caption{}
        \label{subfig:pref30a}
    \end{subfigure}
    \hfill
    \begin{subfigure}{0.325\textwidth}
        \includegraphics[width=\linewidth]{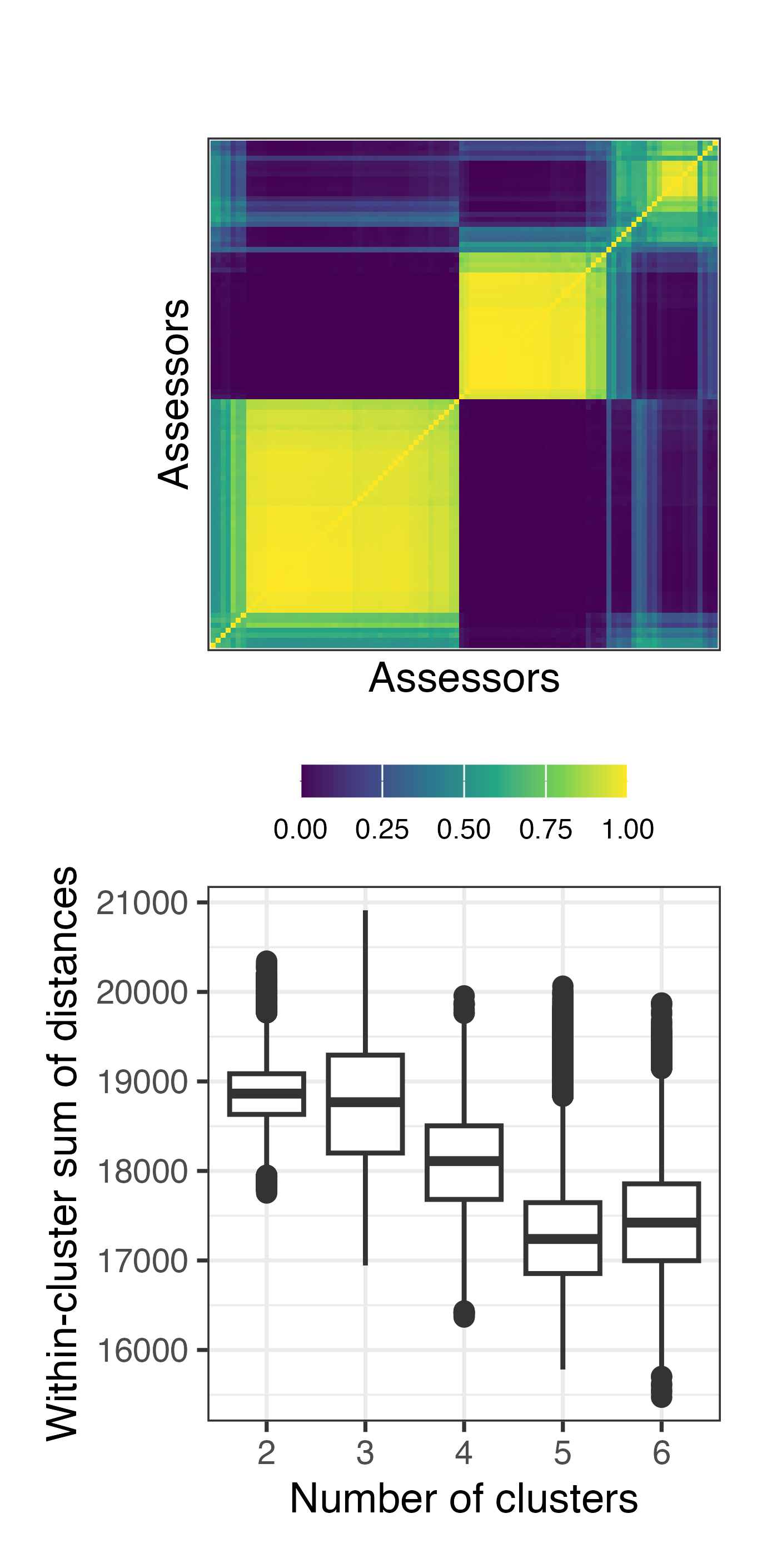}
        \caption{}
        \label{subfig:pref30b}
    \end{subfigure}
    \hfill
    \begin{subfigure}{0.325\textwidth}
        \includegraphics[width=\linewidth]{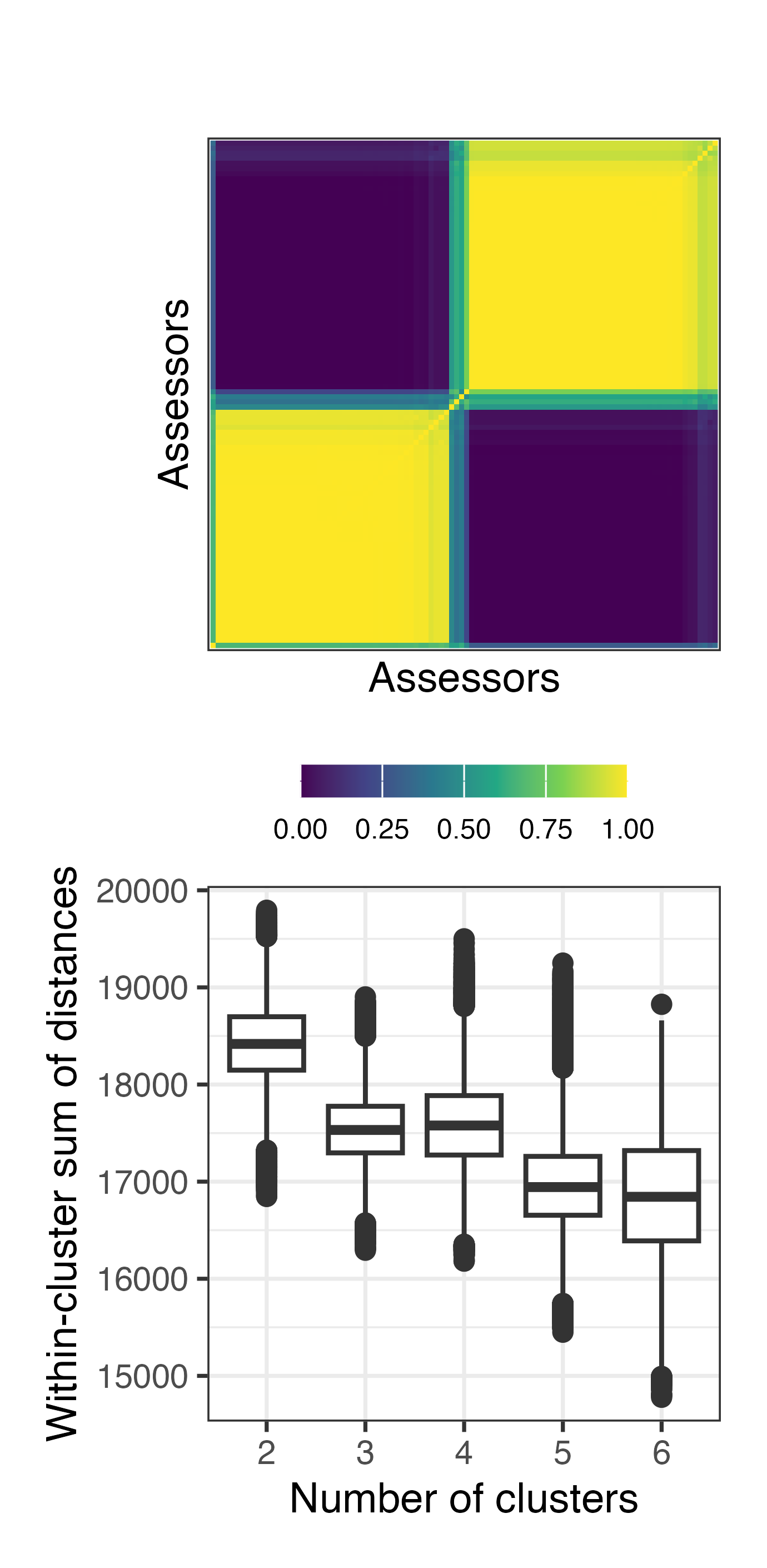}
        \caption{}
        \label{subfig:pref30c}
    \end{subfigure}
    \caption{Co-clustering matrices from DPM3 and box plots of within-cluster sums of distances from finite mixture models for three representative datasets under the \textit{pref30} scenario: (a) DPM3 correctly identifies 3 clusters, while the finite mixture model suggests 4; (b) DPM3 correctly identifies 3 clusters, whereas no clear elbow emerges for the finite mixture model; (c) DPM3 selects 2 clusters, while the finite mixture model correctly identifies 3 clusters.}
    \label{fig:pref30}
\end{figure}

Figure \ref{fig:pref30} shows the results for three representative cases, among the 30 \textit{pref30} datasets. In particular, Figure \ref{fig:pref30}(a) shows an example in which the co-clustering matrix computed from the DPM3 clustering assignments  correctly identifies 3 clusters, whereas the within-cluster sum of distance from the finite mixture models  suggests 4 clusters. Figure \ref{fig:pref30}(b) displays an indecisive elbow plot, ultimately inconclusive, while the nonparametric mixture accurately identifies the presence of 3 clusters. Finally, Figure \ref{fig:pref30}(c) emphasizes an elbow at 3 groups, while the corresponding co-clustering matrix erroneously identifies only 2 clusters.

Table~\ref{tab:resultssimulation} summarizes the results across all 60 experiments, reporting whether each method correctly identified the number of clusters used to generate the data. For finite mixture models, we additionally introduce an \textit{uncertain} category to account for cases in which the elbow plot does not lead to a clear decision. Notice that in the infinite mixture model, such ambiguity is avoided since inference is based on a point estimate of the partition. Therefore, the first advantage of the DPM3 is that the final decision is mathematically determined, eliminating all subjective opinions that arise from the examination of the box-plots.

\begin{table}
\centering
\caption{Accuracy in identifying the number of clusters across 30 simulated datasets for the \textit{top8} and \textit{pref30} scenarios. For DPM3 the number of clusters is inferred from the co-clustering matrix; for finite mixture models (FM) it is selected via the elbow criterion, where \textit{uncertain} denotes the absence of a clear elbow. DPM3 achieves a higher rate of correct identification overall.
}
\label{tab:resultssimulation}
\begin{tabular}{llccc}
    \toprule
    & & \multicolumn{3}{c}{\textbf{Finite Mixture}} \\
    \cmidrule(lr){3-5}
    \textbf{Scenario} & \textbf{DPM3} & Correct & Incorrect & Uncertain \\
    \midrule
    \multirow{2}{*}{\textit{top8}}
        & Correct   & 16 & 6  & 6 \\
        & Incorrect &  1 & 1  & 0 \\
    \midrule
    \multirow{2}{*}{\textit{pref30}}
        & Correct   &  8 & 10 & 8 \\
        & Incorrect &  2 &  1 & 1 \\
    \bottomrule
\end{tabular}
\end{table}

Overall, the results show that the proposed model achieves a higher accuracy in both top-$k$ rankings and pairwise preferences scenarios. In particular, it consistently outperforms the existing approach by correctly identifying three clusters in 93\% and 87\% of cases respectively, compared to 57\% and 33\% for the finite mixture method. Furthermore, only a small number of simulations resulted in a correct identification by the finite mixture model and an incorrect one by DPM3, namely $1$ out of $30$ in the \textit{top8} scenario and $2$ out of $30$ in the \textit{pref30} scenario. The differences seem to be more pronounced in the \textit{pref30} scenario, probably due to the higher level of sparsity induced by this sampling scheme. Moreover, it should be noted that the elbow plot leads to inconclusive results in approximately 25\% of the experiments.

The co-clustering matrices in Figure~\ref{fig:pref30}, obtained via DPM3, show that even when the number of sub-populations is correctly identified, certain assessors may be inaccurately positioned, causing the estimated partition to diverge from the true group structure. In our experience, such inaccuracies occur regardless of the methodology employed and are primarily driven by the high sparsity in the preferences of the misplaced assessors. To assess how much is lost by not knowing the true number of clusters, we compare the partitions recovered by DPM3 and by a finite mixture model with the number of components fixed to the true value of three, across 30 simulated datasets for each scenario. We use the Adjusted Rand Index \citep{hubert1985comparing} (ARI) as a measure of similarity between the estimated and true partition, with values ranging from 0 (random agreement) to 1 (perfect recovery). DPM3 correctly identified three clusters in 26 out of 30 datasets for \textit{pref30} and 28 out of 30 for \textit{top8}. Restricting the attention to these cases, the two methods perform virtually identically: the average ARI with respect to the true partition is 0.792 versus 0.791 for \textit{pref30}, and 0.853 versus 0.850 for \textit{top8}, for DPM3 and the finite mixture model respectively, with DPM3 yielding a partition closer to the truth in 12 out of 26 datasets for \textit{pref30} and 12 out of 28 for \textit{top8}, compared to 10 for the finite mixture model in both scenarios. On average, the two methods disagree on the cluster assignment of only 3.5 assessors out of 100 for \textit{pref30} and 2.6 for \textit{top8}, confirming that the additional uncertainty from not knowing the number of clusters translates into a negligible loss in partition recovery.
\begin{figure}[t]
    \centering
    
    \begin{subfigure}{0.325\textwidth}
        \includegraphics[width=\linewidth]{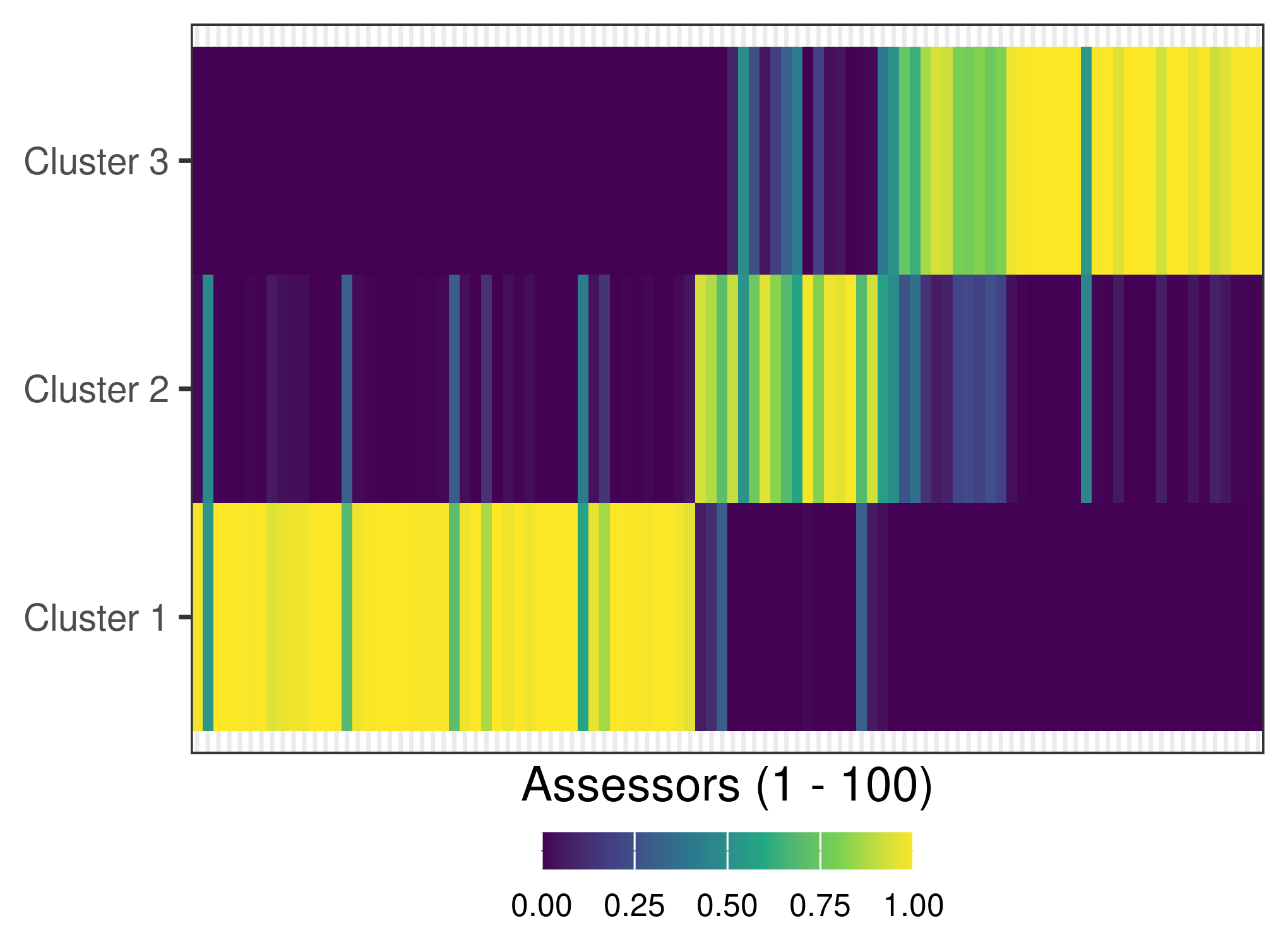}
        \caption{}
        \label{subfig:3clusterstop818}
    \end{subfigure}
    \hfill
    \begin{subfigure}{0.325\textwidth}
        \includegraphics[width=\linewidth]{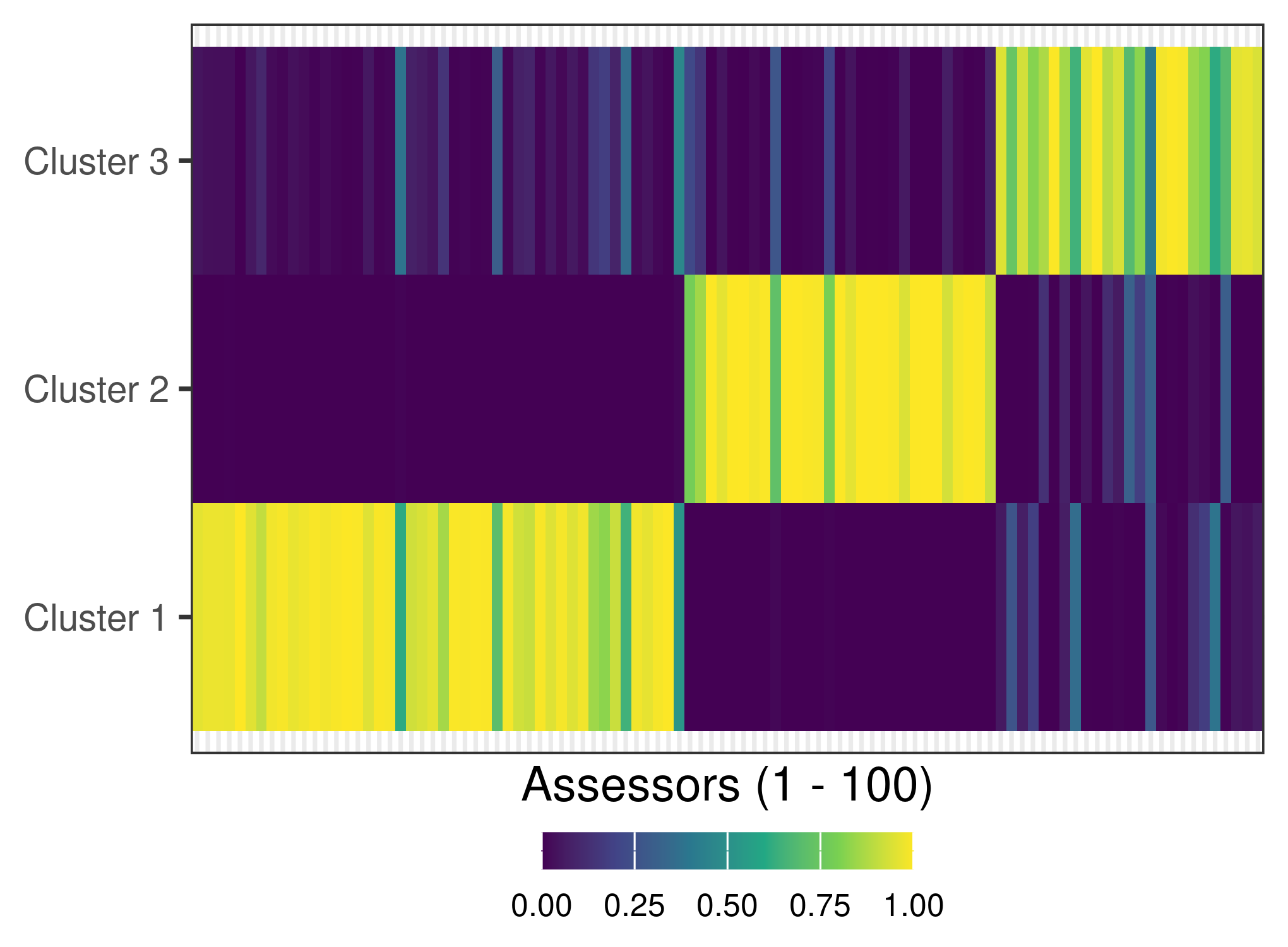}
        \caption{}
        \label{subfig:3clusterspref3018}
    \end{subfigure}
    \hfill
    \begin{subfigure}{0.325\textwidth}
        \includegraphics[width=\linewidth]{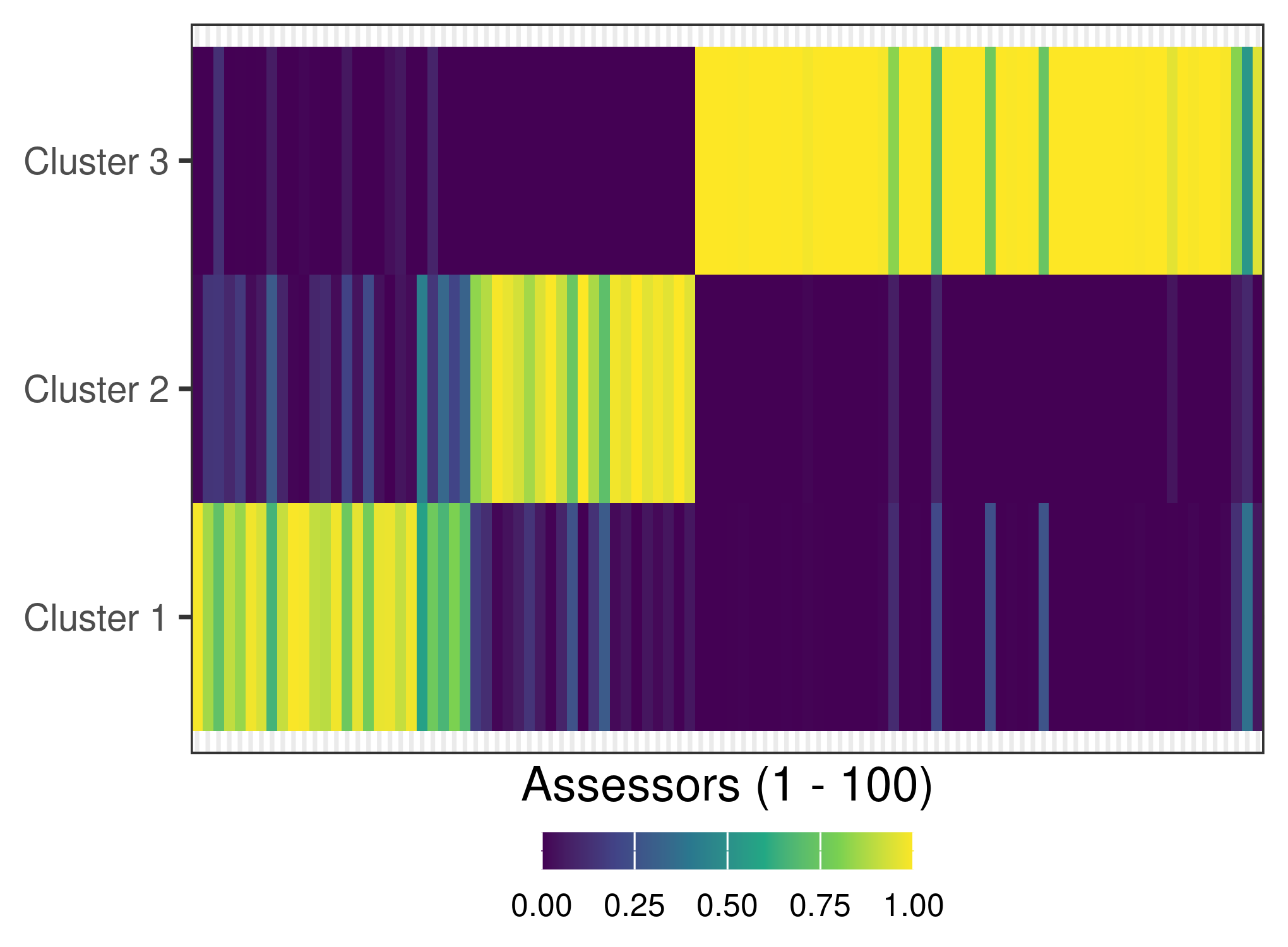}
        \caption{}
        \label{subfig:3clusterspref3020}
    \end{subfigure}
    
    \caption{Cluster assignment probabilities obtained via a finite mixture model with the number of clusters correctly set to 3 for the three datasets whose co-clustering matrices from DPM3 are shown in Figure~\ref{fig:pref30}(a)--(c) respectively.}
    \label{fig:oldmethodmatrix}
\end{figure}

\section{Personalized predictions of movie ratings} \label{sec:movieratings}
In this final study, we illustrate the use of the proposed methodology on a real dataset of movie ratings. In particular, we consider the MovieLens 1M dataset\footnote{www.grouplens.org/datasets/movielens/.}. This comprises ratings ranging from 1 to 5 assigned by 6040 users to a collection of 4000 movies. As in \cite{vitelli2018probabilistic}, we ease calculations by focusing  on the top 200 most rated movies and randomly select 300 assessors. Subsequently, we calculate the Shannon entropy for the 200 most popular movies and retain only the 50 most informative ones. 

Within this refined dataset, we exclude users for whom the random removal of a single rating could result in the absence of any pairwise preferences. More precisely, we do not consider the assessors that provided only one level of rating, and the ones that expressed two levels of ratings with one of them occurring only once. This step allows to remove assessors that provide too limited information, and is essential for being able to test DPM3 as a probabilistic recommender system. Indeed, for each assessor $j$ we randomly sample two movies with different ratings, $A_{i_j}$ and $A_{k_j}$, among those assessed by assessor $j$, and save in a separate set of deleted preferences $\mathcal{D}$ the corresponding pairwise preference , i.e., $A_{i_j} \prec_j A_{k_j}$ if the rating of $A_{i_j}$ is higher than the rating of $A_{k_j}$, while $A_{k_j} \prec_j A_{i_j}$ otherwise. We then randomly select $A_{i_j}$ or $A_{k_j}$ and delete the rating that assessor $j$ assigned to it from the training dataset. Finally, we retrieve all pairwise comparisons implied by the thus obtained dataset.
This process guarantees the self-consistency and transitivity of the derived comparisons. Additionally, it ensures that the preferences deleted in $\mathcal{D}$ cannot be directly inferred from the other ratings remaining in the training dataset, i.e., that no deleted preference in $\mathcal{D}$ will re-appear in the training dataset when computing the training preferences' transitive closure. The final trimmed dataset consists of $N = 206$ assessors and $n = 50$ items.

\begin{figure}[t]
    \centering
    \begin{subfigure}[t]{0.64\textwidth}
        \includegraphics[width=\linewidth]{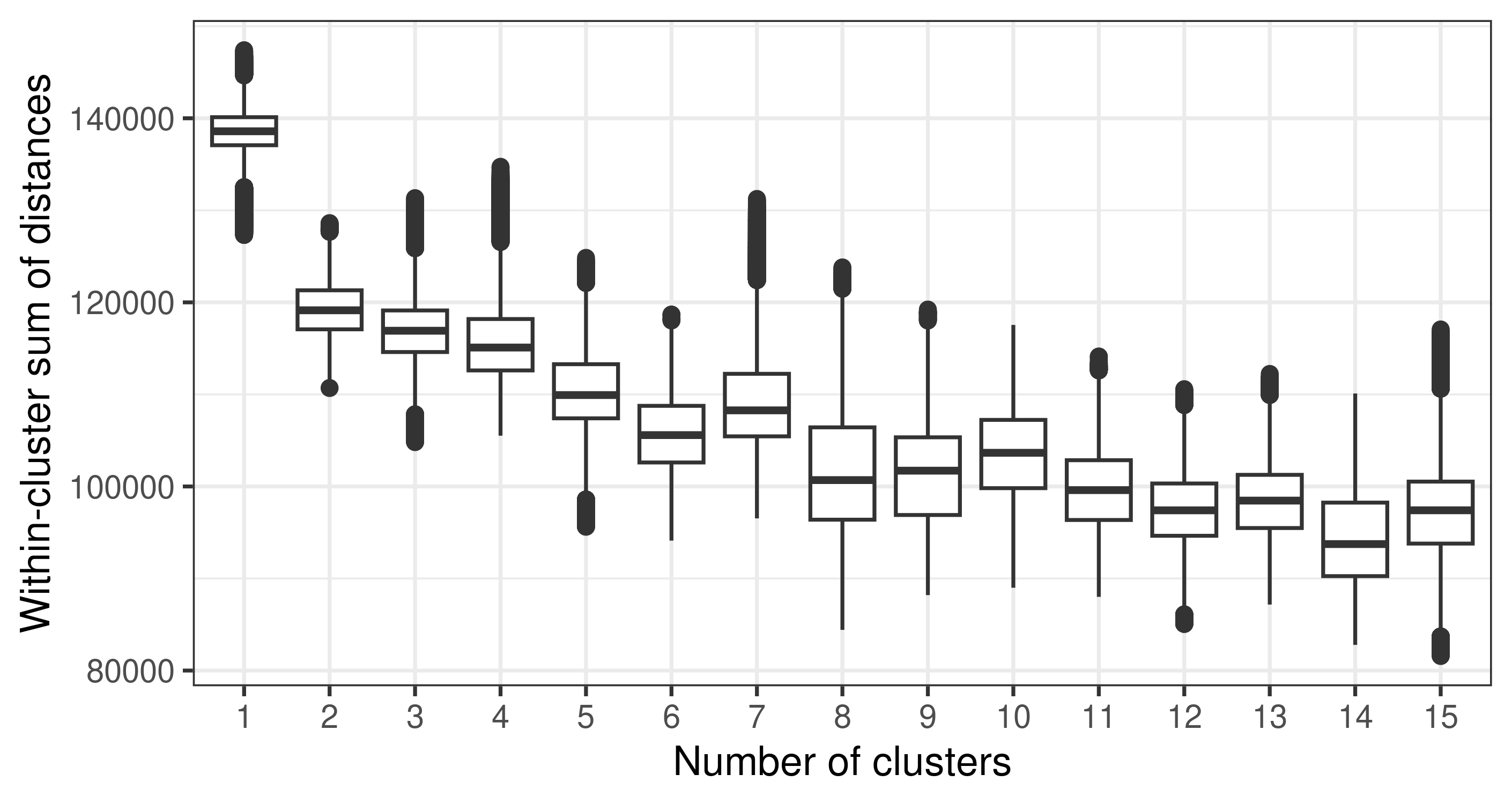}
        \caption{Elbow plot.}
        \label{fig:elbow_movie}
    \end{subfigure}
    \hfill
    \begin{subfigure}[t]{0.35\textwidth}
        \includegraphics[width=\linewidth]{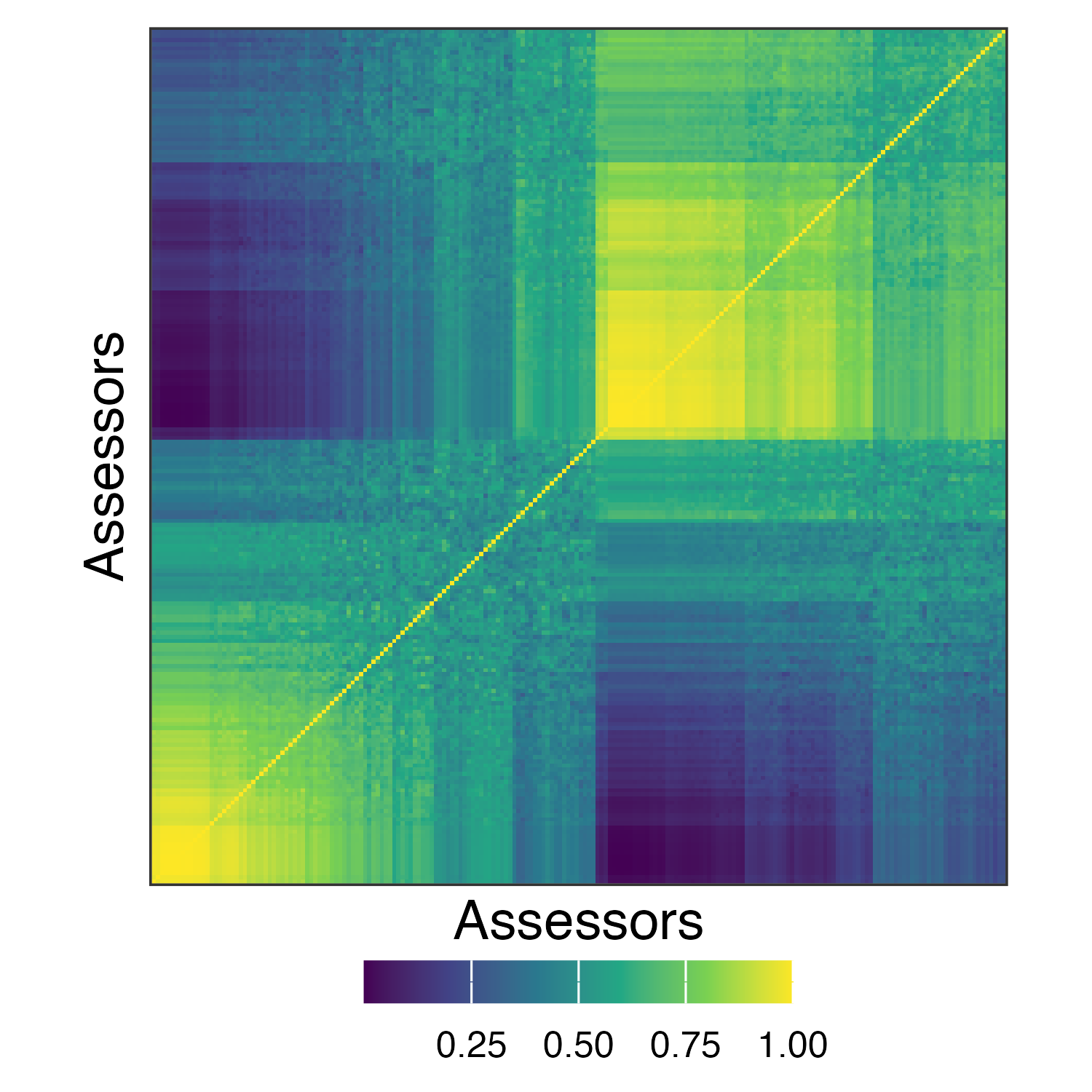}
        \caption{Co-clustering matrix.}
        \label{fig:coclust_movie}
    \end{subfigure}
    \caption{Elbow plot from finite mixtures of Mallows models and co-clustering matrix from DPM3, both fitted to the MovieLens dataset. The finite mixture considers up to 15 clusters.}
    \label{fig:elbow_coclust_movie}
\end{figure}

We conduct an analysis of the trimmed MovieLens dataset by employing the DPM3 algorithm for incomplete data, whose implementation is described in Algorithm~\ref{eq:algorithm2}. We use both a finite and an infinite mixture model, and compare their performance within the primary objective of evaluating the posterior probability of accurately predicting the missing pairwise preferences. For both methods, we choose the footrule distance metric and set the variances of the proposal distributions to $\sigma_{\alpha}=1$ and $L=25$, respectively. This choice reduces the acceptance rates to around 10\%, but allows at the same time a better exploration of the space in a situation of high sparsity and uncertainty. For the finite mixture model, we fit the model with a number of clusters ranging from 1 to 15, in each case for $3\cdot 10^5$ iterations. After discarding one-third of the iterations as burn-in, we generate posterior distributions for the within-cluster sum of distances for each value of $C$, resulting in the boxplots depicted in Figure \ref{fig:elbow_coclust_movie}(a). The determination of the optimal number of clusters appears ambiguous, as choices of $C=2,6,8$ all seem to be reasonable options. For fitting DPM3, we run the MCMC for $2\times 10^5$ iterations and hyperparameter $\psi$ fixed at $7\cdot 10^{-3}$. The number of occupied clusters rapidly converges to 2, as evident from the trace plots in Figure~S4(a) in the Supplementary Material. At convergence, the posterior distribution of the number of non-empty clusters is concentrated at 2, with a posterior probability of around 98\%. Trace plots for the scale parameters and empirical cluster probabilities of the five most persistent labels throughout the chain also clearly show convergence of the chain towards a structure with two primary sub-populations, with one sub-population being slightly larger than the other (see Figures~S4(b) and~S4(c) in the Supplementary Material). After setting the burn-in at 25,000 iterations, we also compute the co-clustering matrix, presented in Figure \ref{fig:elbow_coclust_movie}(b). This also clearly shows a two-blocks structure. 

\begin{figure}[t]
    \centering
    \begin{subfigure}[t]{0.35\textwidth}
        \includegraphics[width=\linewidth]{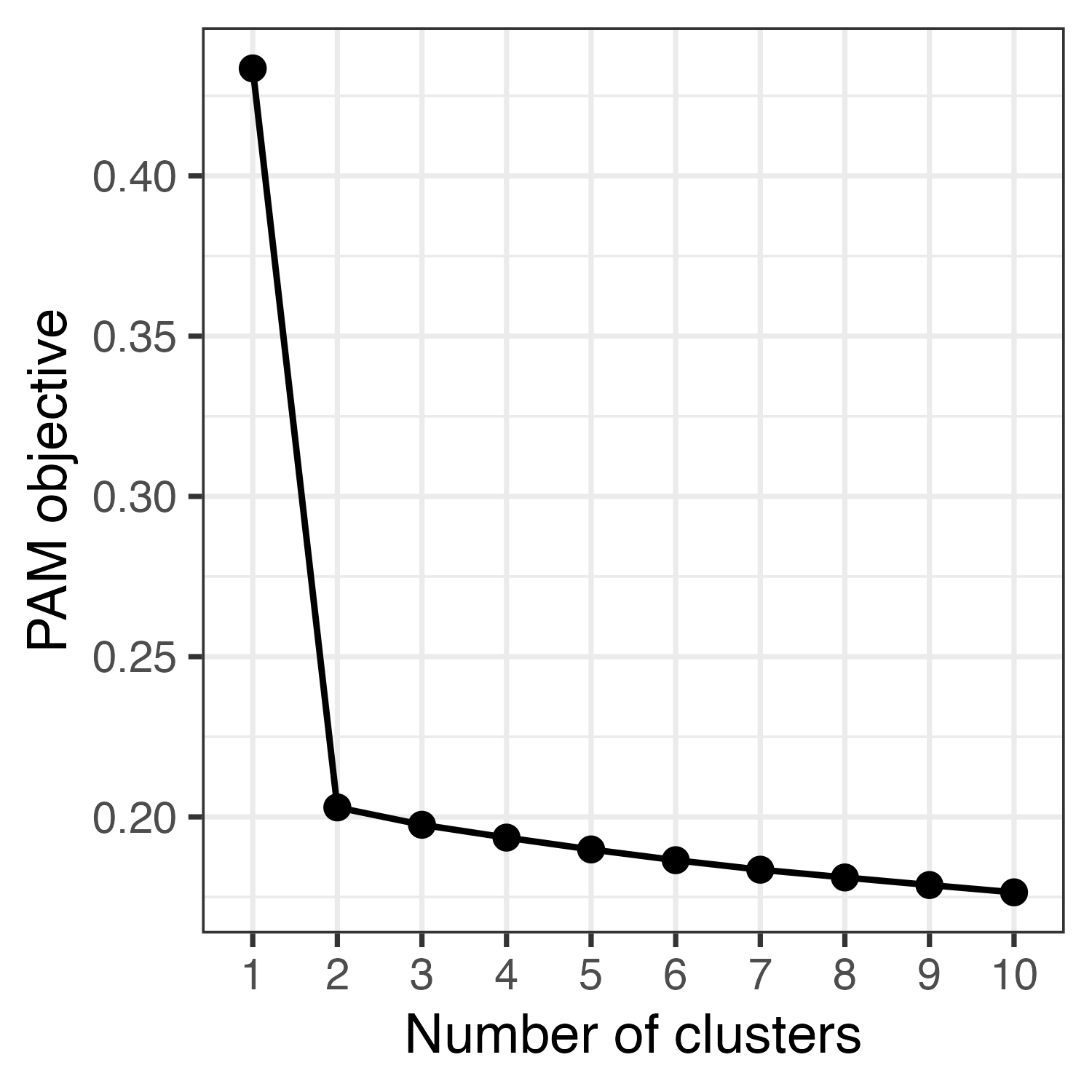}
        \caption{PAM elbow plot.}
        \label{fig:elbow_pam}
    \end{subfigure}
    \hfill
    \begin{subfigure}[t]{0.64\textwidth}
        \includegraphics[width=\linewidth]{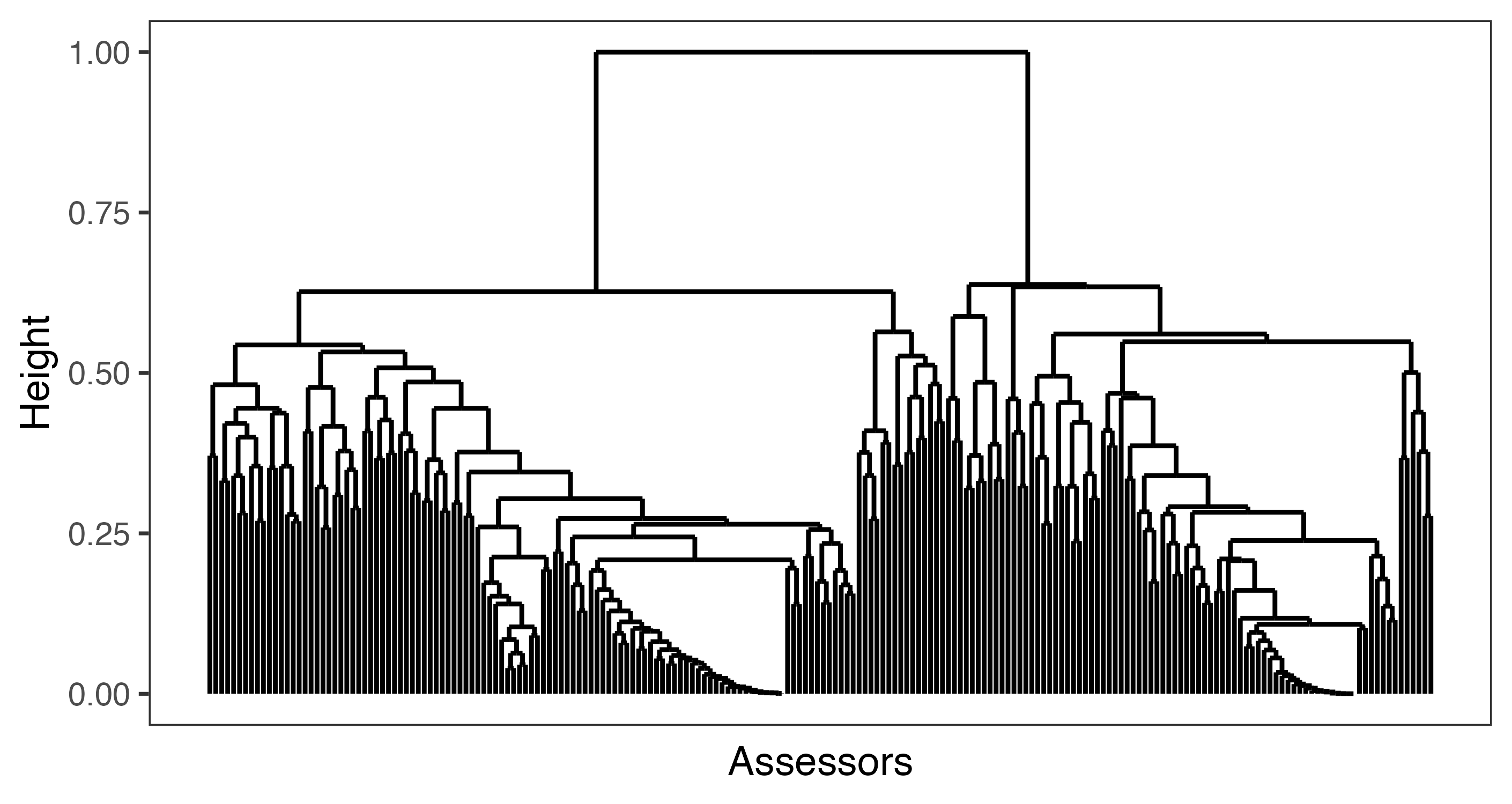}
        \caption{Hierarchical clustering dendrogram.}
        \label{fig:dendrogram}
    \end{subfigure}
    \caption{PAM elbow plot and hierarchical clustering applied to the co-clustering matrix of the MovieLens dataset of Figure~\ref{fig:coclust_movie}. Both analyses suggest the presence of two clusters.}
    \label{fig:clustering}
\end{figure}

Despite the clear indication of two clusters in all the plots presented so far, the minimization of the lower bound of the posterior expected variation of information using the \texttt{minVI} function \citep{WadeBayesian} is not sufficiently sensitive to detect a meaningful two groups partition among assessors. Hence, we post-process the co-clustering matrix with classical clustering methods, specifically hierarchical clustering and PAM (Partitioning Around Medoids) \citep{kaufman2009finding}. These results are shown in Figure~\ref{fig:clustering}: both the dendrogram and the PAM elbow plot consistently suggest a partition into two clusters. The average silhouette width shown in Figure~S5 of the Supplementary Material for both PAM (Figure~S5(a)) and hierarchical clustering (Figure~S5(b)) further confirms the presence of two distinct clusters. Since the partitions induced by the two methods differ in the assignment of only 8 assessors out of 206, in what follows we arbitrarily concentrate on the partition detected by hierarchical clustering.

In particular, we examine the covariates associated with the users in order to draw conclusions on the composition of the two sub-populations. Cluster 1 comprises 124 assessors with an average age of 32.1, while the remaining 82 assessors in Cluster 2 have an average age of 26.0. Compared to the overall average age of 29.7, Cluster 1 is slightly larger and consists, on average, of older individuals. Regarding gender composition, Cluster 1 exhibits a higher proportion of female assessors (approximately 30\%) compared to Cluster 2 (approximately 12\%). The age disparity becomes evident when examining the occupation of the assessors. Figure~S6 in the Supplementary Material shows the four most represented occupations among assessors, revealing further differences in the composition of the two c with the most significant differences in prevalence between the two clusters, along with their respective percentages within the clusters and across the entire dataset. As expected, Cluster 2 exhibits a higher presence of college and graduate students and a lower representation of executive and managerial positions. Additionally, a distinction is observed in the categories of lawyers and academic/education positions, with a higher prevalence of the first category in Cluster 1 and of the second category in Cluster 2.

\begin{figure}[t]
    \centering
    \begin{subfigure}{0.49\textwidth}
        \includegraphics[width=\linewidth]{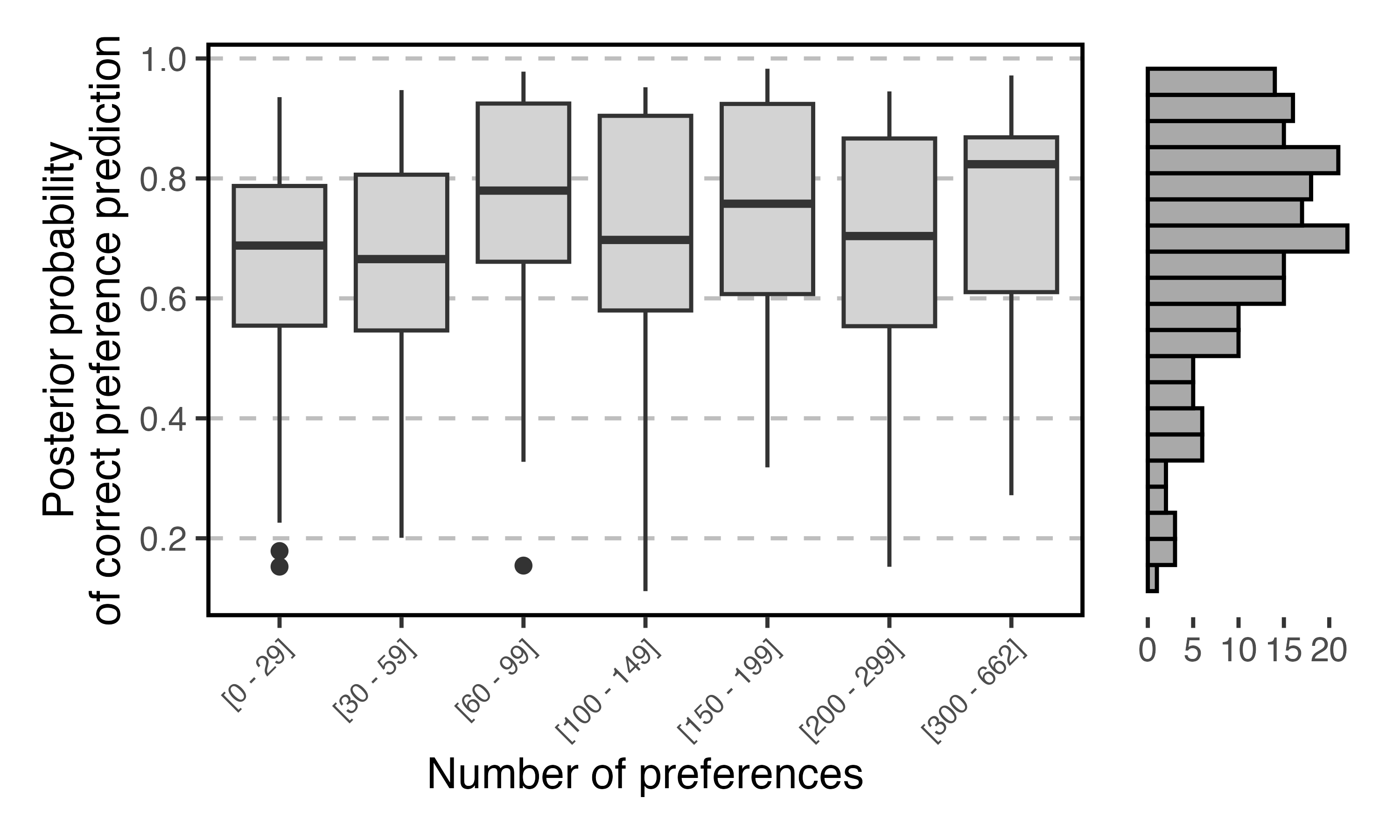}
        \caption{Finite mixture model with two clusters.}
        \label{fig:prob_old}
    \end{subfigure}
    \hfill
    \begin{subfigure}{0.49\textwidth}
        \includegraphics[width=\linewidth]{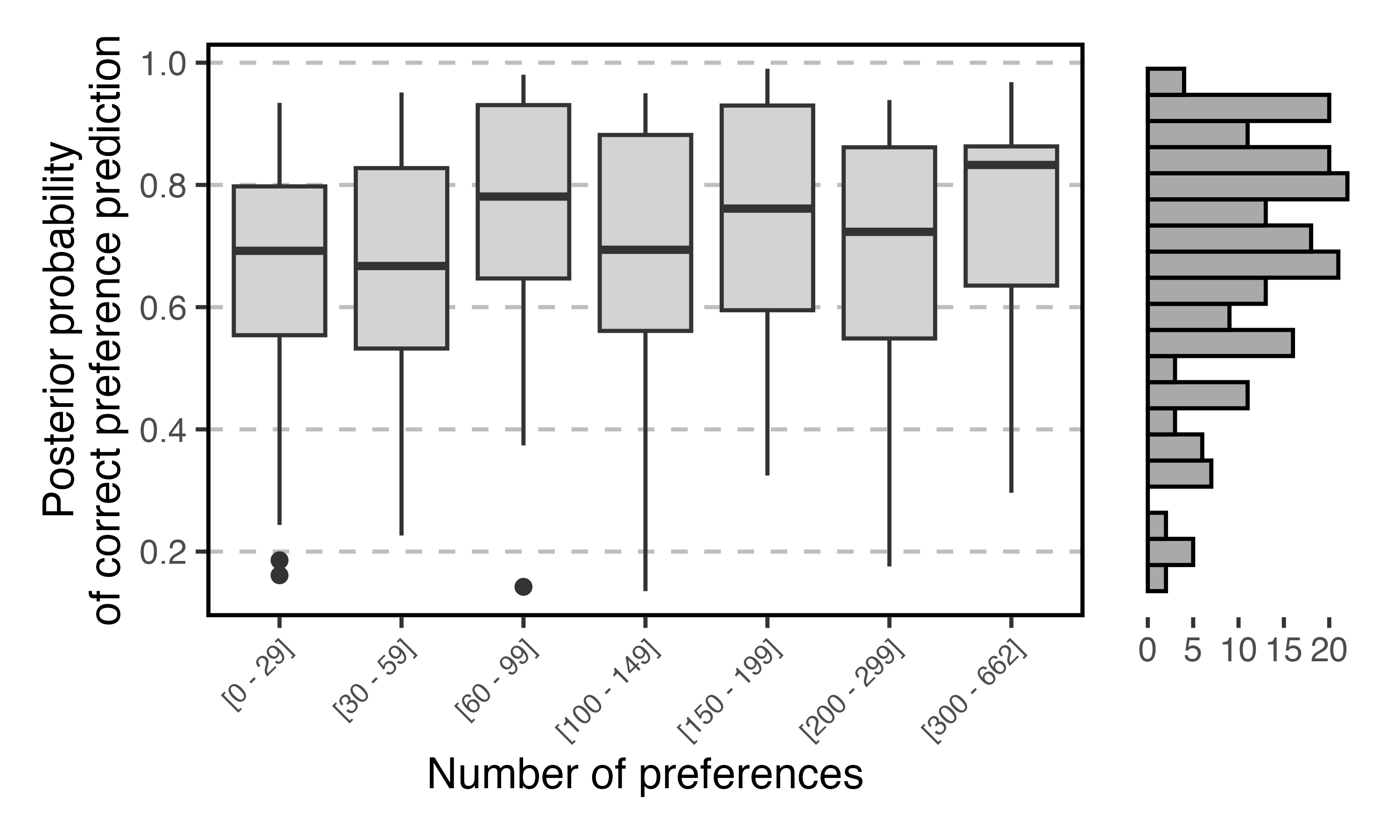}
        \caption{Dirichlet process mixture model.}
        \label{fig:prob_new}
    \end{subfigure}
    \caption{Posterior probability of correctly predicting held-out pairwise preferences for the finite mixture model with two clusters and the Dirichlet process mixture model, as a function of the number of preferences provided by each assessor.}
    \label{fig:prob}
\end{figure}

After evaluating the number of occupied clusters in the DPM3 solution for the trimmed MovieLens dataset, we now proceed to the evaluation of prediction performance. Specifically, we compare the finite mixture model with two components and the Dirichlet process mixture model. For both models, we analyze the posterior distribution of augmented rankings for individual assessors and calculate the posterior probability of accurately predicting the omitted preference initially removed from the dataset. This is achieved by computing the proportion of augmented rankings sampled in the posterior, for each assessor, in which the correct ordering between the two involved movies is retrieved. Figures \ref{fig:prob}(a) and \ref{fig:prob}(b) show the boxplots of the posterior probability of accurately predicting the omitted preference across assessors, for the finite mixture and DPM3 models, respectively. For these figures, the assessors have been stratified according to the number of available pairwise comparisons depending on the ratings they have provided (on the x-axis). The histogram on the right-side of both panels provides an overview of the overall distribution of such posterior probabilities across assessors. As expected, there is a slight improvement in the prediction for those assessors who expressed many pairwise judgments. The two models exhibit nearly identical behavior, with a mean probability of 0.68 and a median of 0.7 for the finite mixture model and 0.71 for the infinite mixture, meaning that the posterior probability of the correct pairwise ordering is relatively high. This consistent performance is visually evident in Figure~S7 in the Supplementary Material, where the posterior probabilities of correct prediction are directly compared: the scatter plot concentrates around the main diagonal of the square, denoting an overall equivalence in performance.

In summary, the DPM3 clearly identifies two distinct clusters in the MovieLens dataset, while the finite mixture model leads to a more ambiguous determination of the number of components (Figure~\ref{fig:elbow_coclust_movie}). Despite this difference, the two approaches deliver comparable predictive accuracy and partition recovery, confirming the Bayesian Mallows model as an effective tool for probabilistic recommendations. The DPM3 offers the additional advantage of a more principled determination of the number of clusters, particularly valuable in sparse data settings where the elbow criterion may be inconclusive.

\section{Conclusion}\label{sec:conc}
\markboth{Conclusion}{Conclusion}
The Dirichlet process mixture model provides a versatile approach to Bayesian density estimation, particularly in the case of mixture models. This technique can be effectively applied to the Mallows model for rank data, achieving a balance between flexibility, computational efficiency, and interpretability. The proposed model integrates well within this framework, supporting various right-invariant distances and managing incomplete data.

In their foundational work, \citet{vitelli2018probabilistic} used classical finite mixture models for clustering, employing Gibbs sampling within a MCMC algorithm to update cluster assignments and proportions. In that work, model selection was conducted solely by analyzing posterior distributions of within-cluster distance sums for different sub-population fits. In contrast, we propose using a Metropolis-within-Gibbs sampler that targets the posterior distribution of Dirichlet process mixtures of Mallows models, allowing for the evaluation of the co-clustering matrix post-burn-in to derive data partitions. While \citet{WadeBayesian} advocated for a criterion based on minimizing the lower bound of the posterior expected variation of information, this method showed limited effectiveness under high uncertainty, as demonstrated in the MovieLens experiment. Therefore, we suggest exploring alternative approaches such as PAM and hierarchical clustering, which are well suited for working directly on similarity or distance matrices.

The empirical studies aimed to achieve two key objectives: comparing the effectiveness of finite and infinite mixture methods in accurately determining the number of non-empty clusters in simulated datasets, and assessing their ability to recover missing preferences in real movie rating data. The nonparametric approach outperformed the finite mixture method in identifying the correct number of clusters, showing better accuracy in both top-$k$ and pairwise preference scenarios. These matched closely those derived from the finite mixture model, assuming the number of groups was known. The MovieLens dataset was used to evaluate the infinite mixture model's performance on real user ratings for movies, revealing that the Bayesian nonparametric approach offered superior clarity in identifying the optimal number of clusters while still delivering precise personalized predictions. Deleted comparisons were identified with an average precision of 68\% by both the infinite mixture model and the finite mixture model with two components. Overall, the convergence of parameters was generally satisfactory, despite the sparse nature of the data, even though the finite mixture model showed quite high uncertainty around the decision on the number of groups.

While the Bayesian nonparametric approach provides distinct benefits, it also presents challenges due to the larger parameter space, which sometimes complicates the convergence of the Markov chain to the posterior distribution and requires significant computational resources. Currently, the DPM3 algorithm is less efficient than the finite mixture model already available in \texttt{BayesMallows}. Future efforts will focus on optimizing the MCMC sampler for the infinite mixture model to ensure that the choice between finite and infinite mixture is determined by modeling needs rather than computational limitations. Additionally, 
differently to the case of finite mixtures, managing label switching in infinite mixtures requires not only correcting the swapping of existing labels but also accounting for the creation and disappearance of labels across iterations, making automated solutions an interesting direction for future research.

\bibliographystyle{chicago}
\bibliography{bibliography}

\clearpage
\input{arxiv_supp}

\end{document}

%% file: arxiv_supp.tex
\begin{center}
    {\LARGE \textbf{Supplementary Material}}\\[0.5em]
    {\large Bayesian nonparametric Mallows model for clustering preference data}
\end{center}

\section*{Supplement A. Illustrative example}

In the following we illustrate the use of our software on a randomly generated dataset from a finite mixture of Mallows models. We generated complete rankings for $N=200$ assessors and $n=20$ items using the procedure presented in Appendix C of \citet{vitelli2018probabilistic}, with parameters specified in Table~\ref{tab:setting}.

 To run Algorithm 1, we use the \texttt{compute\_mallows\_dpmixture} function, adjusting the thinning parameters and the $\psi$ value. The default distance is the footrule, but Cayley, Kendall, Hamming, Spearman, and Ulam distances are also supported. Non-specified parameters are set to their default values: $\psi_{\text{init}}=5$, $\lambda = 0.1$, $\sigma_{\alpha}=0.1$, $L=\max(1, \lfloor n/5 \rfloor)$.

\begin{lstlisting}
dpm_test <- compute_mallows_dpmixture(data, psi = 0.01, nmc = 100000,
                clus_thin = 10, rho_thinning = 10, alpha_jump = 10)
\end{lstlisting}

\begin{figure}
    \centering
    \begin{subfigure}{0.32\textwidth}
        \includegraphics[width=\linewidth]{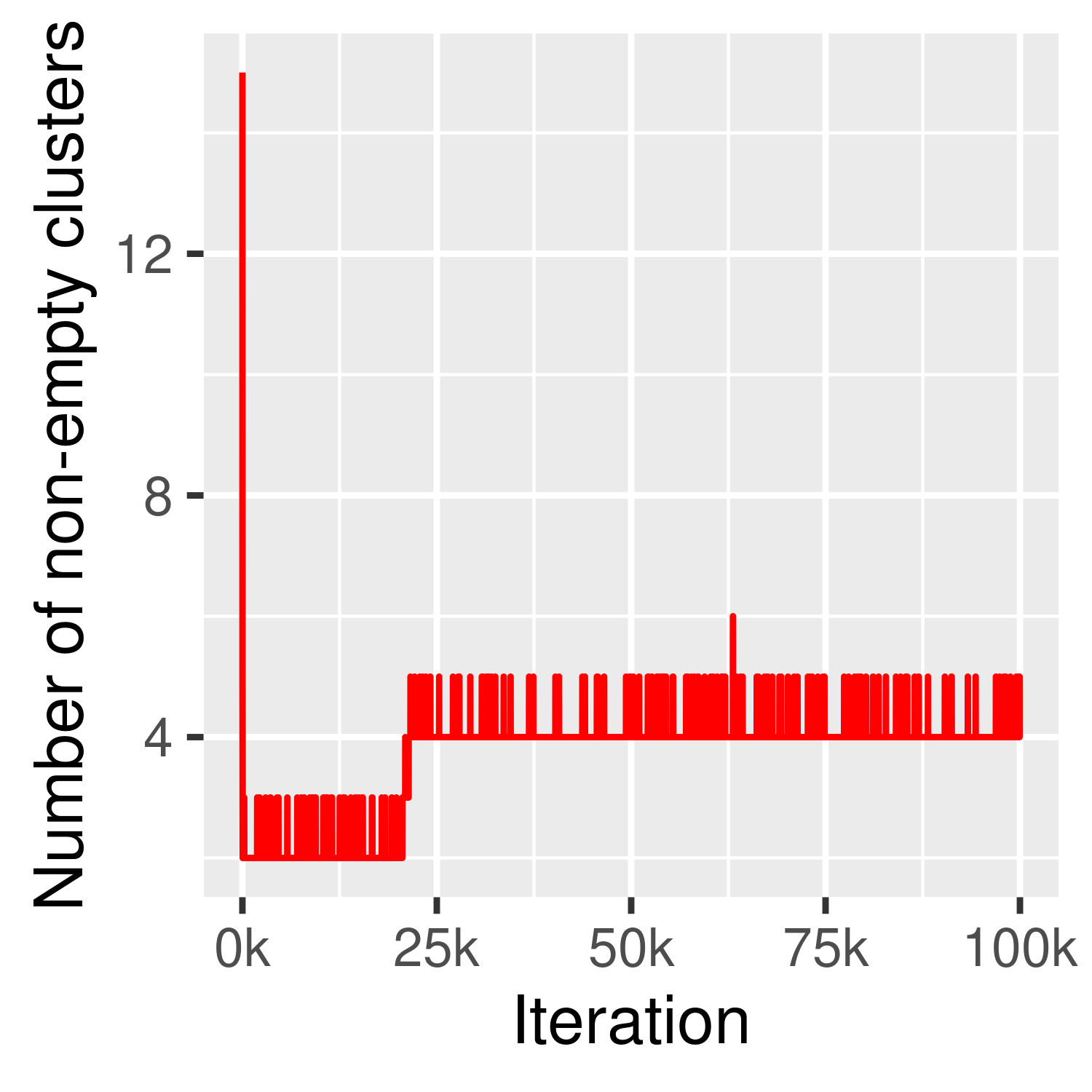}
        \caption{Number of non-empty clusters.}
        \label{fig:trace_nclusters_example}
    \end{subfigure}
    \hfill
    \begin{subfigure}{0.32\textwidth}
        \includegraphics[width=\linewidth]{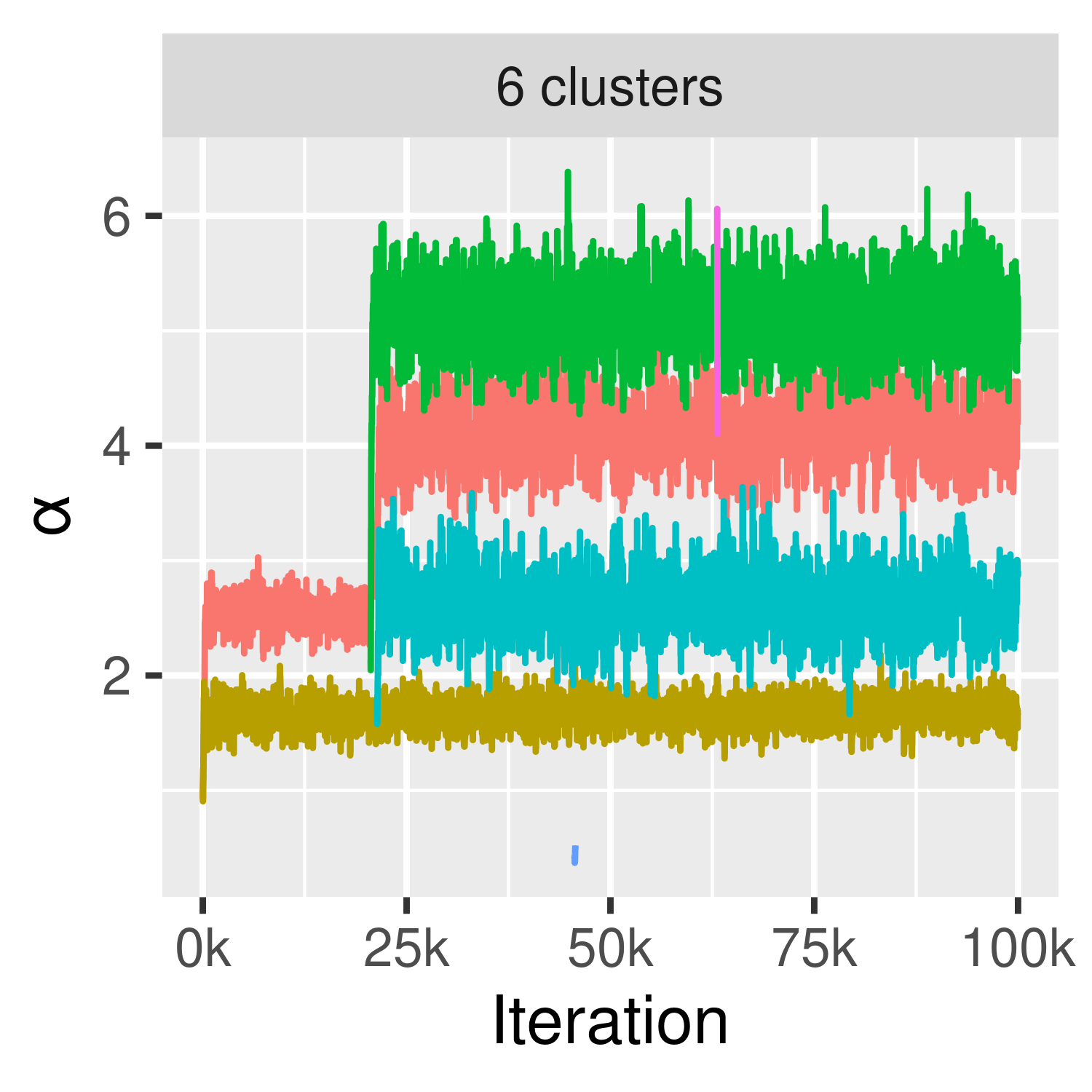}
        \caption{Scale parameter $\alpha$ per cluster.}
        \label{fig:trace_alpha_example}
    \end{subfigure}
    \hfill
    \begin{subfigure}{0.32\textwidth}
        \includegraphics[width=\linewidth]{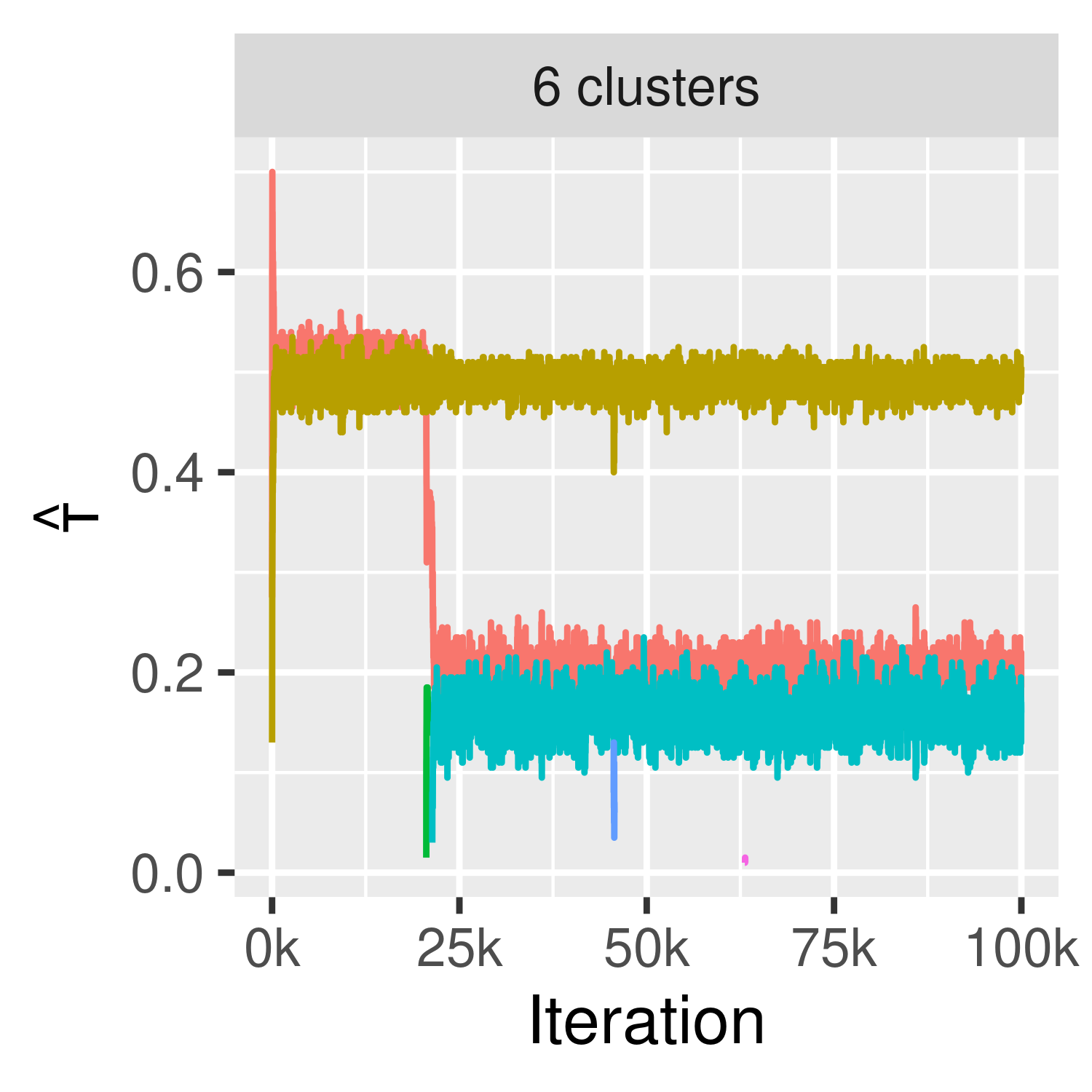}
        \caption{Empirical cluster probabilities $\hat{\tau}$.}
        \label{fig:trace_clusterprobs_example}
    \end{subfigure}
    \caption{Trace plots for the MCMC convergence diagnostics. The six most persistent cluster labels are shown for $\alpha$ and $\hat{\tau}$.}
    \label{fig:trace_example}
\end{figure}

 Trace plots to assess convergence of the chain can be generated using the function \linebreak \texttt{assess\_convergence\_dpmixture}, which by default displays the number of non-empty clusters over iterations (Figure~\ref{fig:trace_example}(a)). The same function can plot either $\alpha$ or the estimated cluster proportions $\hat{\tau}$ by specifying the \texttt{parameter} argument. In these cases, the \texttt{n} argument controls how many clusters to display; the software automatically selects the \texttt{n} most persistent labels, where persistence is defined as the number of samples in which the label appears among the cluster assignments.

\begin{lstlisting}
assess_convergence_dpmixture(dpm_test)
assess_convergence_dpmixture(dpm_test, parameter = "alpha", n = 6)
assess_convergence_dpmixture(dpm_test, parameter = "empirical_cluster_probs", n = 6)
\end{lstlisting}

 The corresponding plots are shown in Figure~\ref{fig:trace_example}. After approximately 25,000 iterations, the chain exhibits satisfactory convergence towards a four-cluster structure. Following an initial burn-in phase in which two predominant clusters emerge, each encompassing roughly half of the assessors, two new labels appear, leading to a stable subdivision into four groups. As visible from Figure~\ref{fig:trace_example}(a), new clusters continue to be generated but none leads to a state with higher posterior probability. This is further confirmed by Figures~\ref{fig:trace_example}(b) and \ref{fig:trace_example}(c), where the 5\textsuperscript{th} and 6\textsuperscript{th} most persistent labels are barely noticeable. 
 
 We therefore set the burn-in and compute the co-clustering matrix using the function \texttt{compute\_co\_clustering}.

\begin{lstlisting}
dpm_test$burnin <- 25000
dpm_test$co_clustering <- compute_co_clustering(dpm_test)
\end{lstlisting}

\begin{figure}
    \centering
    \includegraphics[width=0.5\textwidth]{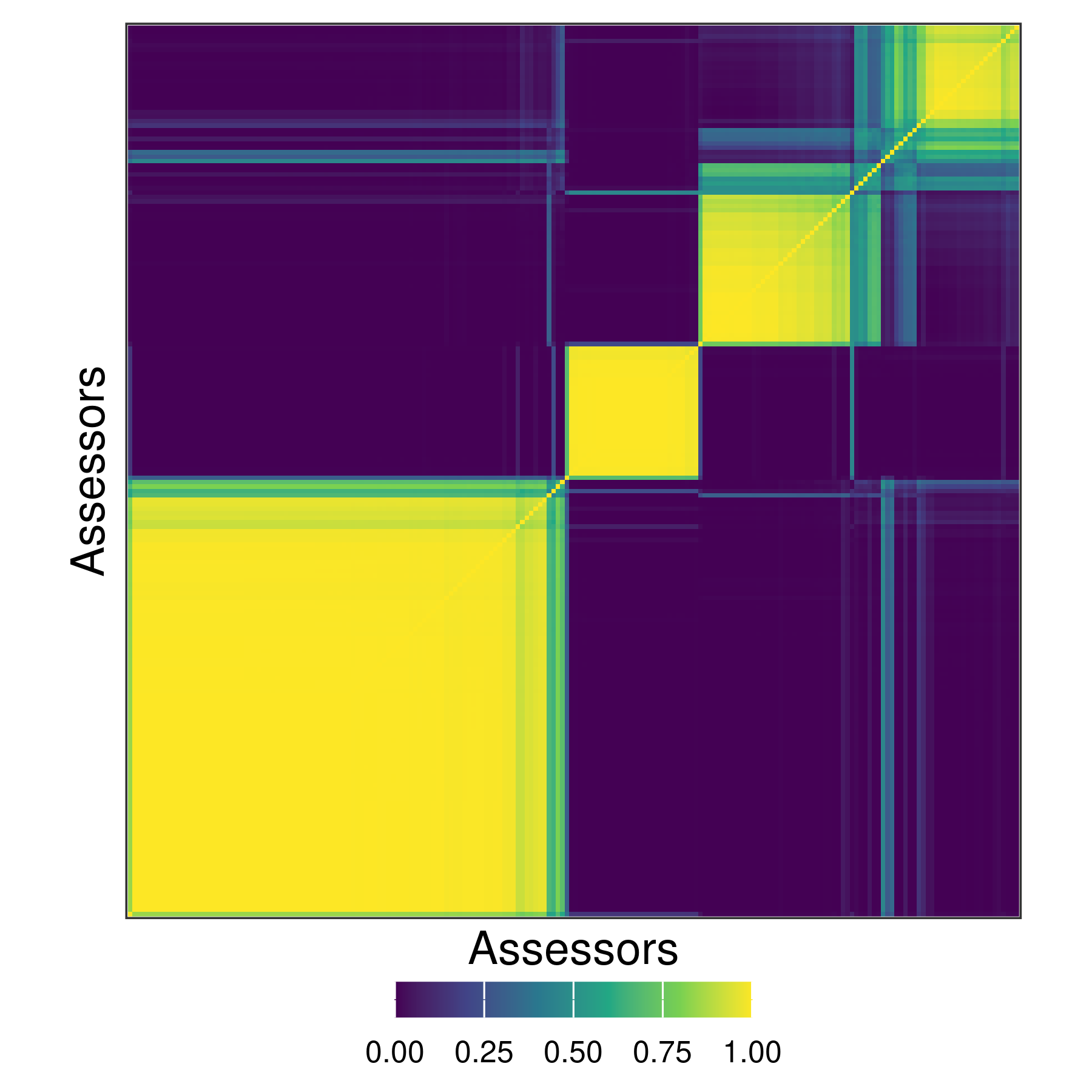}
    \caption{Co-clustering matrix with assessors reordered according to the estimated partition, clearly revealing four well-separated blocks.}
    \label{fig:co_clust_example}
\end{figure}

When reordering the assessors, the co-clustering matrix appears as in Figure~\ref{fig:co_clust_example}, clearly revealing four well-separated blocks. The partition can then be estimated using the function \texttt{partition\_estimate}, which by default relies on the \texttt{avg} method of the \texttt{minVI} function from the \texttt{mcclust.ext} package \citep{WadeBayesian, mcclustext}. The cluster labels are stored in the \texttt{cl} field of the \texttt{partition} attribute.

\begin{lstlisting}
dpm_test$partition <- partition_estimate(dpm_test)
max(dpm_test$partition$cl)
#> [1] 4
table(dpm_test$partition$cl)
#>  1  2  3  4
#> 98 31 41 30
\end{lstlisting}

\begin{figure}
    \centering
    \begin{subfigure}{0.49\textwidth}
        \includegraphics[width=\linewidth]{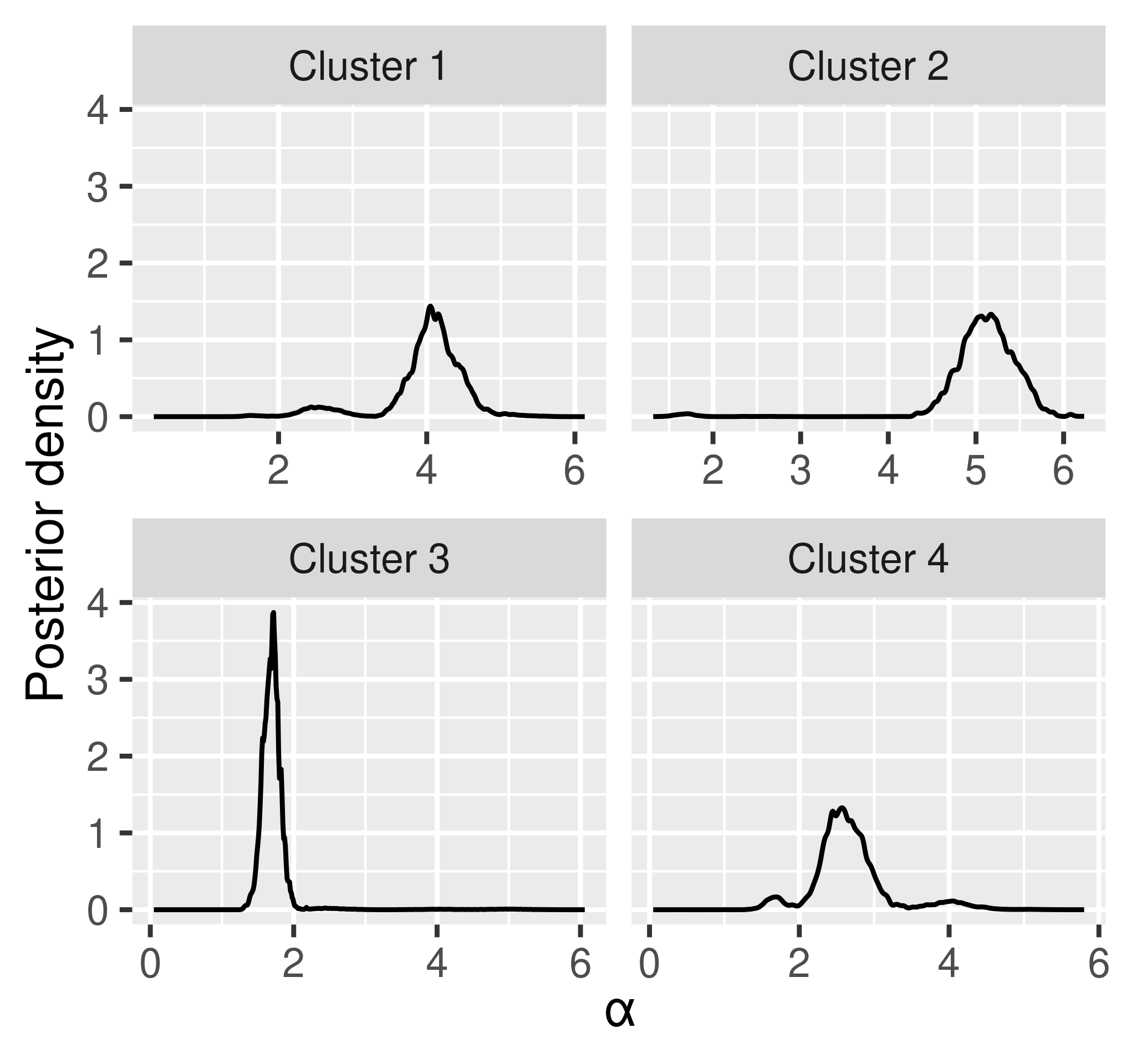}
        \caption{Posterior distribution of $\alpha^*_c$, $c=1,\dots,4$.}
        \label{fig:alpha_postexample}
    \end{subfigure}
    \hfill
    \begin{subfigure}{0.49\textwidth}
        \includegraphics[width=\linewidth]{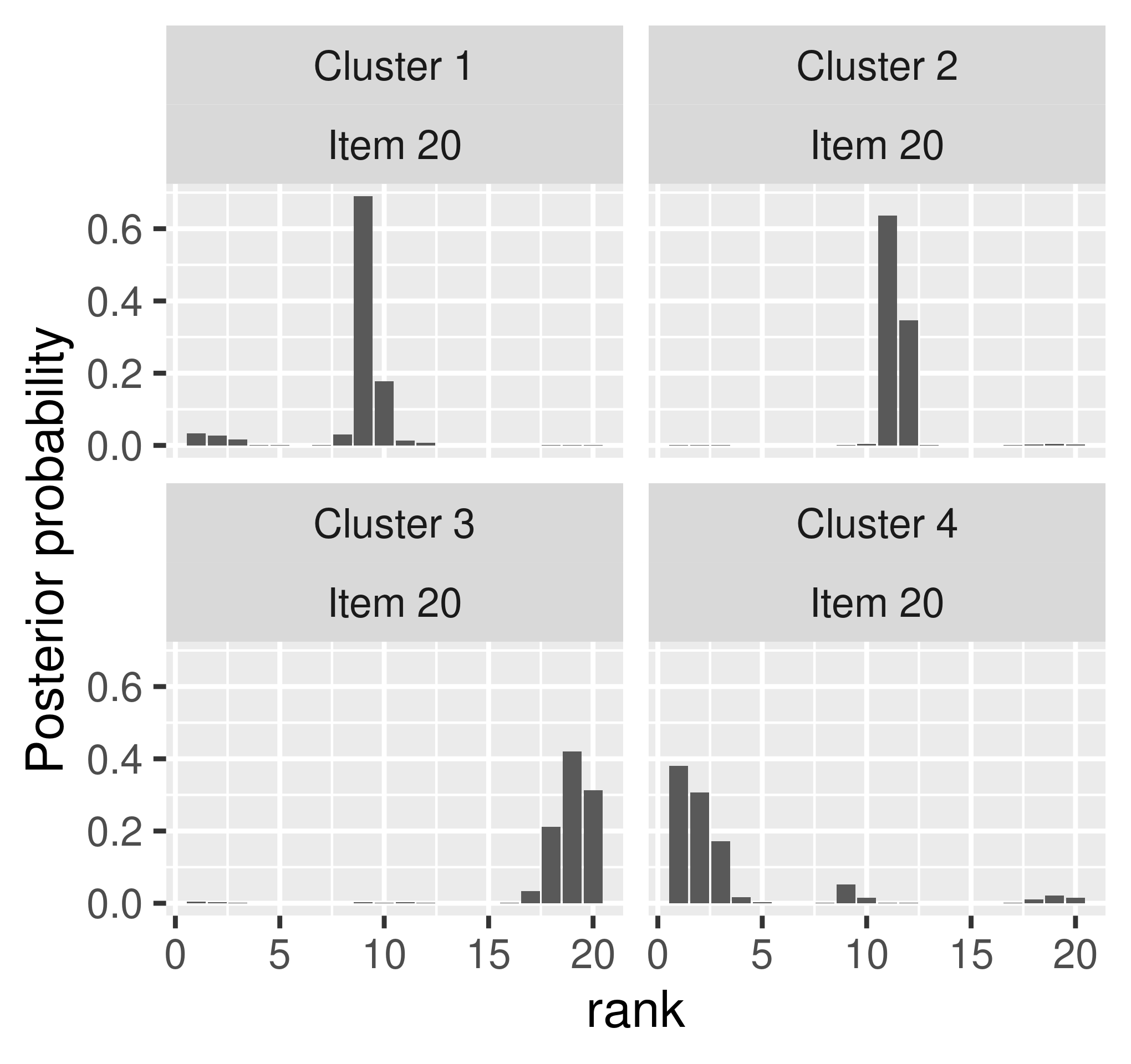}
        \caption{Posterior distribution of $\rho^*_{c,20}$, $c=1,\dots,4$.}
        \label{fig:rho_plotexample}
    \end{subfigure}
    \caption{Posterior distributions of the cluster-specific parameters $\alpha$ and $\rho$ conditioned on the estimated partition.}
    \label{fig:posteriors_example}
\end{figure}

 The four clusters are accurately identified, with only occasional misclassifications among assessors. The co-clustering matrix reveals some uncertainty in placing these assessors, reflecting the nuanced nature of their classification. Conditioning on the estimated partition, posterior distributions of the cluster-specific parameters can be visualized via the \texttt{plot} function, as shown in Figures~\ref{fig:posteriors_example}(a) and~\ref{fig:posteriors_example}(b).

\begin{lstlisting}
plot(dpm_test, parameter = "alpha")
plot(dpm_test, parameter = "rho", items = 20)
\end{lstlisting}

Credible intervals and consensus rankings can additionally be obtained using the functions \texttt{compute\_posterior\_intervals} and \texttt{compute\_consensus}, both inherited from the original \texttt{BayesMallows} package \citep{BayesMallowsRjournal}.

\begin{lstlisting}
compute_posterior_intervals(dpm_test, parameter = "alpha")
compute_consensus(dpm_test, type = "CP")
\end{lstlisting}

 \begin{table}[H]
    \centering
    \caption{Cluster-specific consensus rankings, precision parameters, mixing weights, and resulting cluster sizes used in data generation.}
    \label{tab:setting}
    \begin{tabular}{lcccc}
        \toprule
        Cluster & $\boldsymbol{\rho}$ & $\alpha$ & $\tau$ & Size \\
        \midrule
        1 & $(1,\dots,20)$                & 2 & 0.5 & 98 \\
        2 & $(20,\dots,1)$                & 3 & 0.2 & 36 \\
        3 & $(20,\dots,11,1,\dots,10)$    & 4 & 0.2 & 37 \\
        4 & random in $\mathcal{P}_{20}$  & 5 & 0.1 & 29 \\
        \bottomrule
    \end{tabular}
\end{table}

\clearpage
\section*{Supplement B. Additional plots for the MovieLens application}

\begin{figure}[H]
    \centering
    \begin{subfigure}{0.32\textwidth}
        \includegraphics[width=\linewidth]{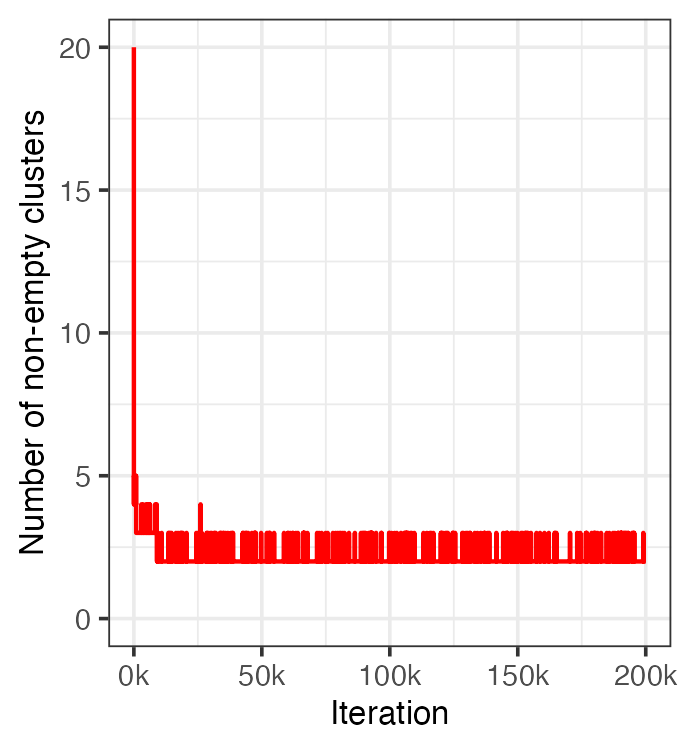}
        \caption{Number of non-empty clusters.}
        \label{fig:trace_nclusters}
    \end{subfigure}
    \hfill
    \begin{subfigure}{0.32\textwidth}
        \includegraphics[width=\linewidth]{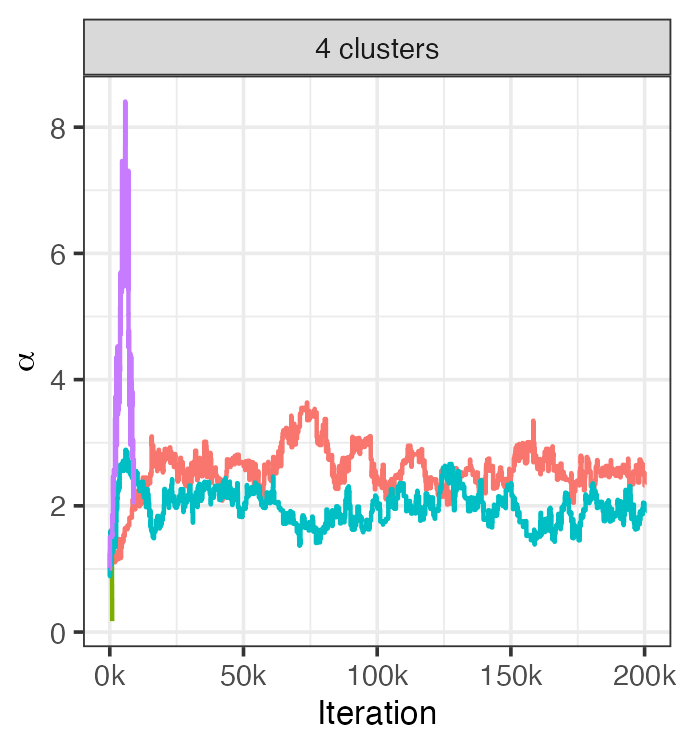}
        \caption{Scale parameter $\alpha$ per cluster.}
        \label{fig:trace_alpha}
    \end{subfigure}
    \hfill
    \begin{subfigure}{0.32\textwidth}
        \includegraphics[width=\linewidth]{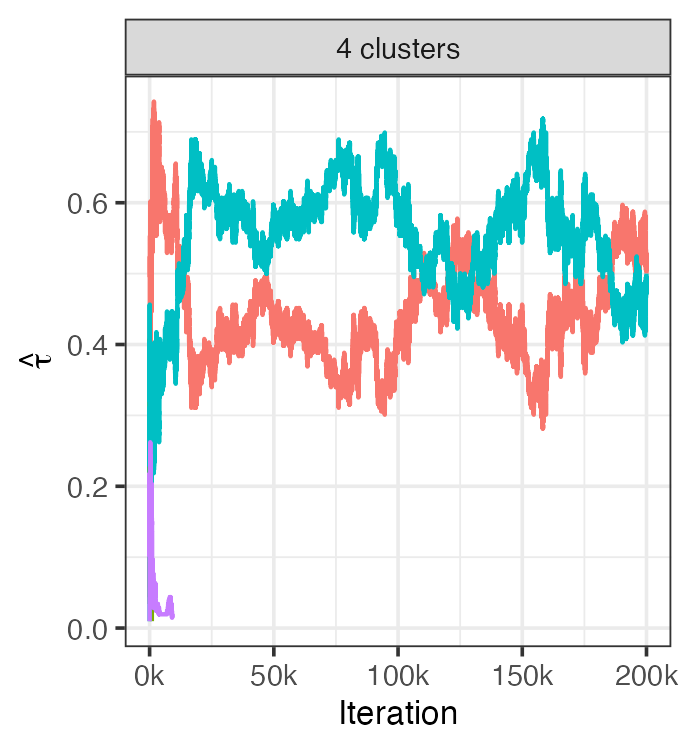}
        \caption{Empirical cluster probabilities $\hat{\tau}$.}
        \label{fig:trace_clusterprobs}
    \end{subfigure}
    \caption{Trace plots for the MCMC convergence diagnostics of the DPM3 model fitted to the MovieLens dataset.}
    \label{fig:trace_all}
\end{figure}

\begin{figure}[H]
    \centering
    \begin{subfigure}{0.48\textwidth}
        \includegraphics[width=\linewidth]{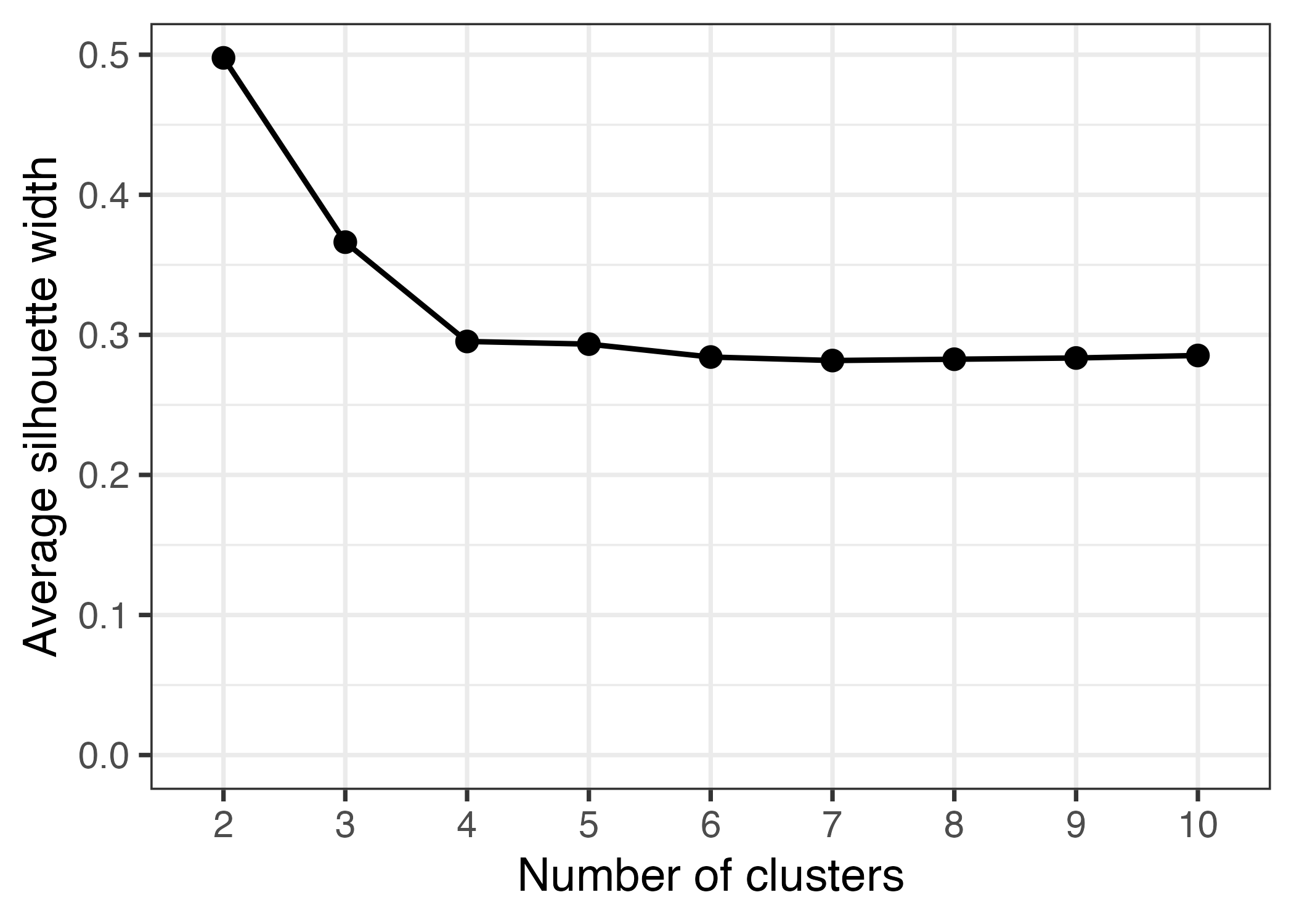}
        \caption{PAM silhouette.}
        \label{fig:silhouette_pam}
    \end{subfigure}
    \hfill
    \begin{subfigure}{0.48\textwidth}
        \includegraphics[width=\linewidth]{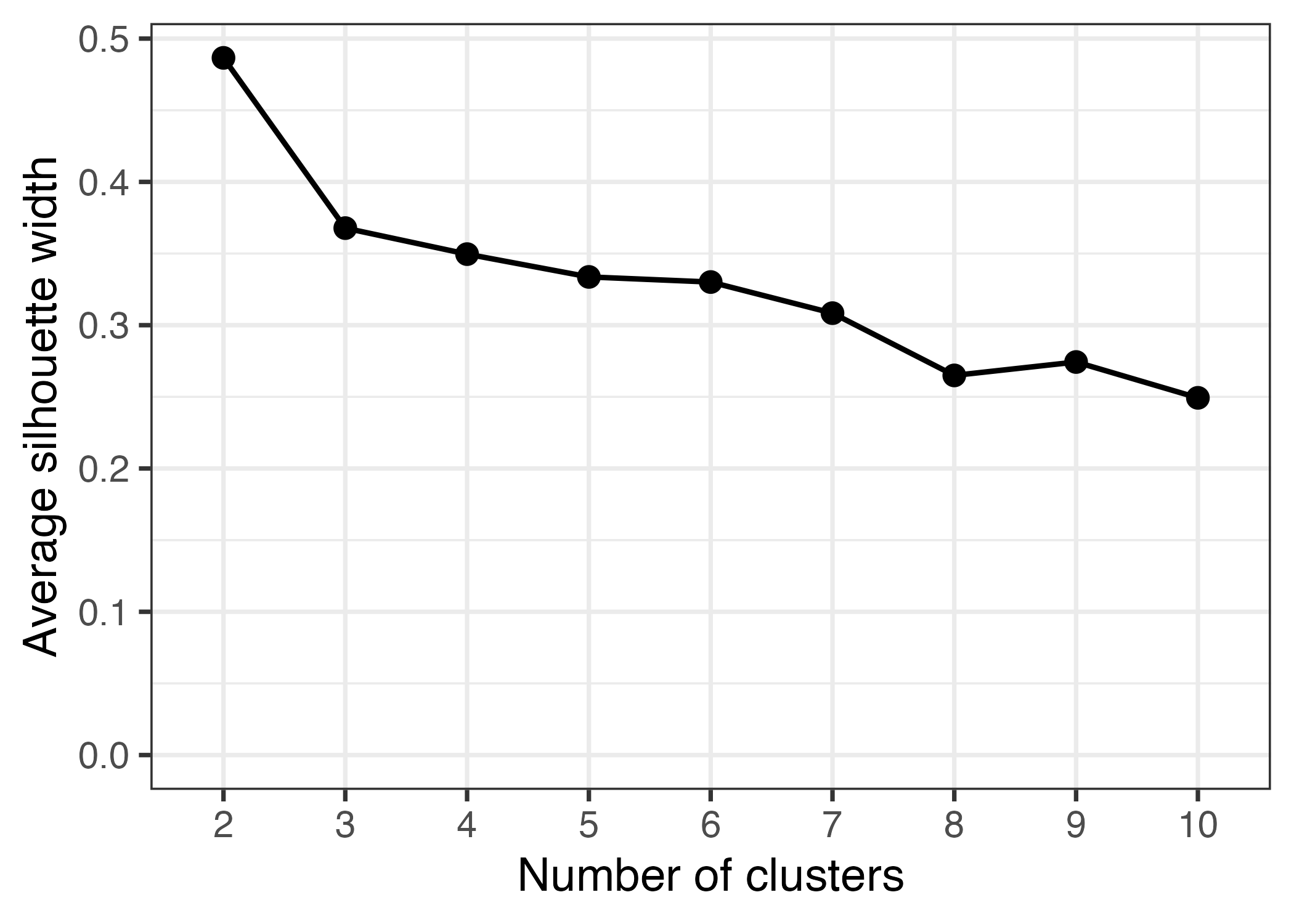}
        \caption{Hierarchical clustering silhouette.}
        \label{fig:silhouette_hclust}
    \end{subfigure}
    \caption{Average silhouette width as a function of the number of clusters for PAM and hierarchical clustering applied to the co-clustering matrix of the MovieLens dataset.}
    \label{fig:silhouette_supplement}
\end{figure}

\begin{figure}[t]
    \centering
    \includegraphics[width=0.65\textwidth]{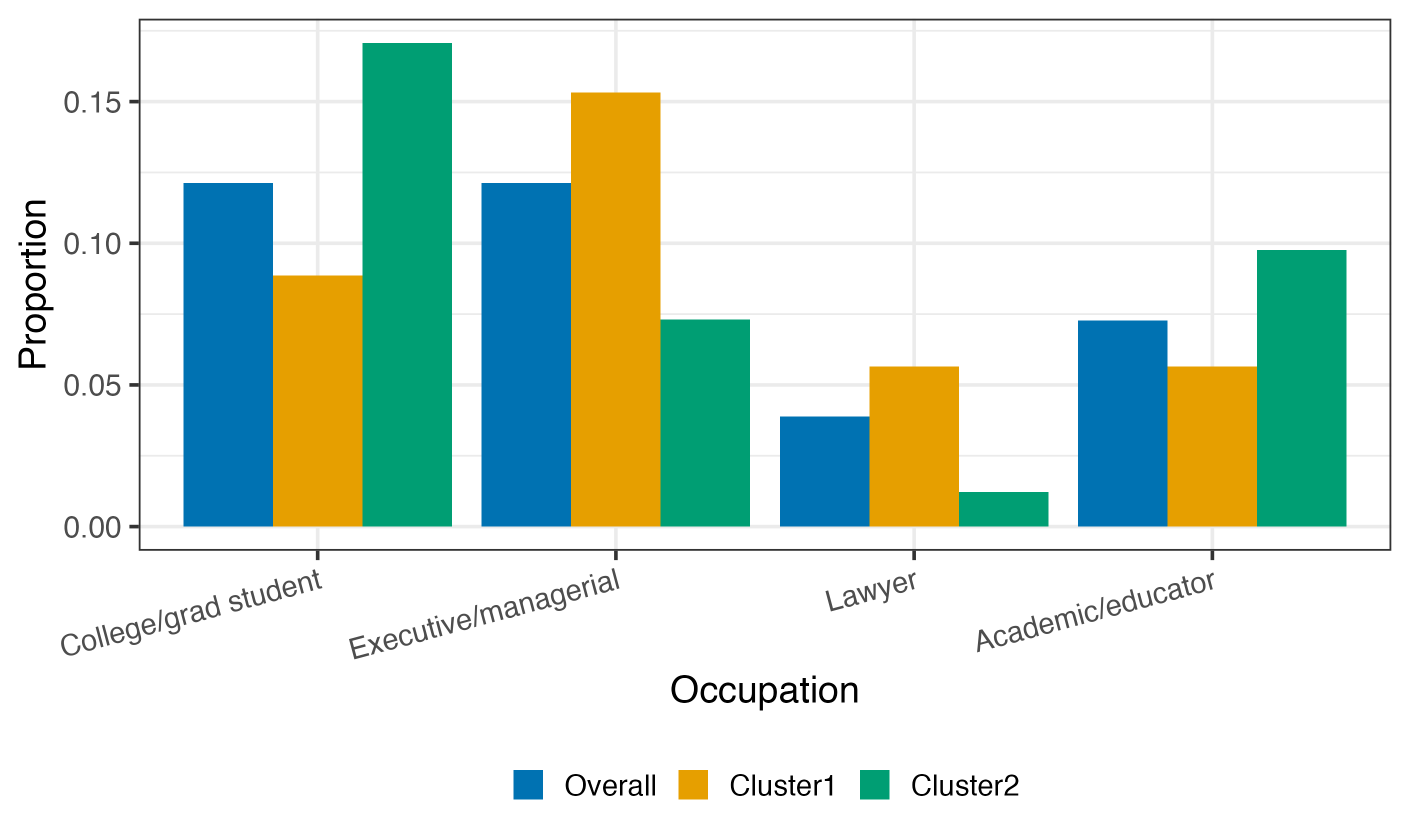}
    \caption{Proportion of assessors per occupation category, compared across the two clusters identified by hierarchical clustering and the overall population, for the MovieLens dataset.}
    \label{fig:occupation}
\end{figure}

\begin{figure}[t]
    \centering
    \includegraphics[width=0.5\textwidth]{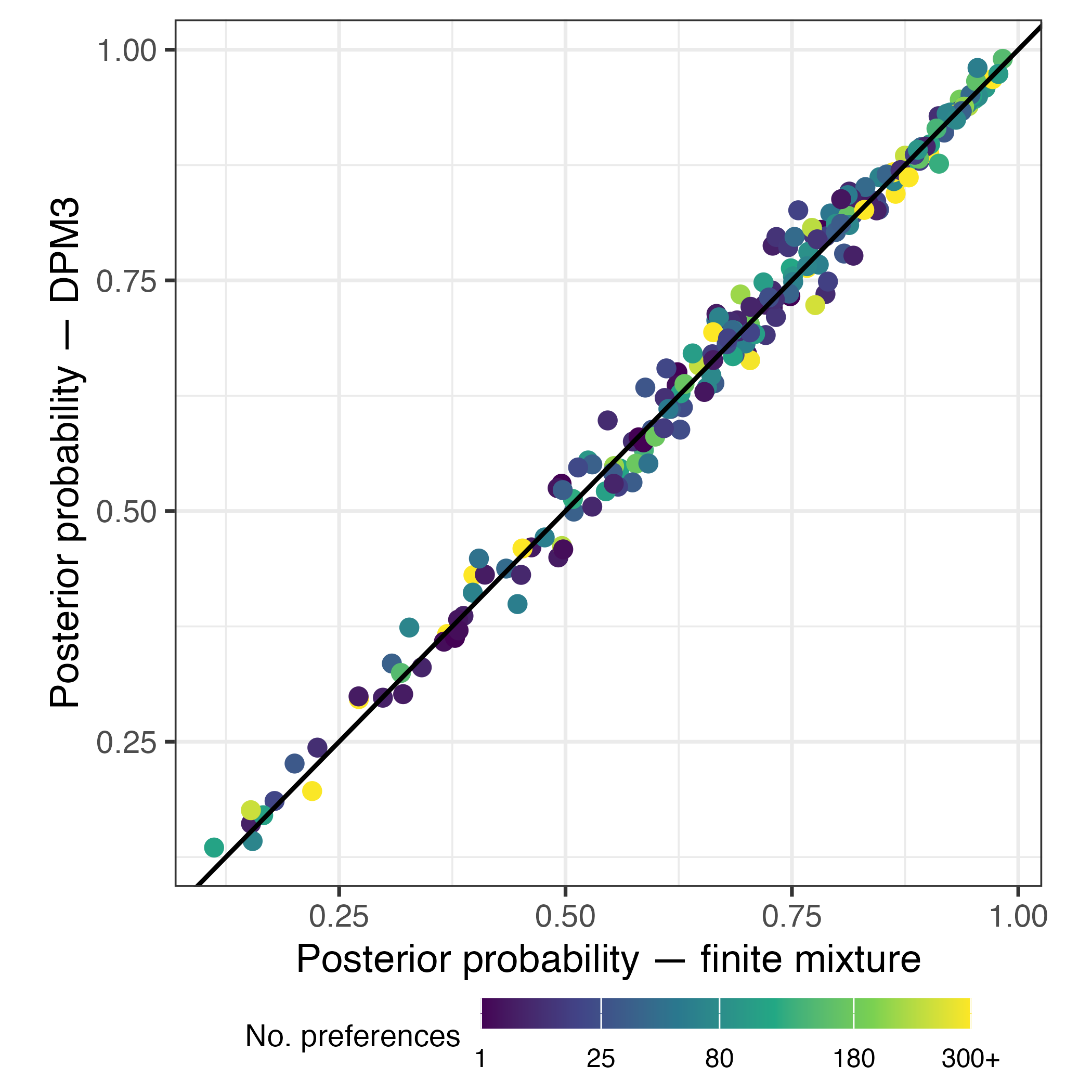}
    \caption{Posterior probabilities of correctly predicting held-out pairwise preferences, comparing the finite mixture model with two clusters against DPM3. Each point represents one assessor, colored by the number of preferences derived from their ratings.}
    \label{fig:diag_plot}
\end{figure}

This supplement provides additional diagnostic and exploratory evidence supporting the MovieLens analysis in Section~4. Figure~\ref{fig:trace_all} shows MCMC trace plots for key DPM3 parameters, indicating good mixing and convergence. Figure~\ref{fig:silhouette_supplement} reports average silhouette widths for PAM and hierarchical clustering over a range of cluster numbers, with both methods consistently favoring a two-cluster partition of the co-clustering matrix. Figure~\ref{fig:occupation} compares occupation distributions across the two inferred clusters and the full population, revealing systematic differences in their professional composition. Figure~\ref{fig:diag_plot} evaluates held-out predictive performance for pairwise preferences, contrasting the finite two-cluster mixture with DPM3 and highlighting assessor-level variation linked to rating density.

